\documentclass[12pt]{iopart}

\usepackage{mathabx}
\DeclareMathSymbol{\fdia}{2}{mathb}{"0C}

\usepackage{iopams}
\usepackage{dsfont}
\usepackage{graphics}

\newcommand{\ket}[1]{\left|{#1}\right\rangle}
\newcommand{\bra}[1]{\left\langle{#1}\right|}
\newcommand{\braket}[2]{\langle{#1}|{#2}\rangle}

\begin{document}

\title{Fermion quadrature bases for Wigner functionals}

\author{Filippus S. Roux}
\address{University of Kwazulu-Natal, Private Bag X54001, Durban 4000, South Africa}
\ead{stefroux@gmail.com}

\begin{abstract}
A Grassmann functional phase space is formulated for the definition of fermionic Wigner functionals by identifying suitable fermionic operators that are analogues to boson quadrature operators. Instead of the Majorana operators, we use operators that are defined with relative spin transformations between the ladder operators. The eigenstates of these operators are shown to provide orthogonal bases, provided that the dual space is defined with the incorporation of a spin transformation. These bases then serve as quadrature bases in terms of which Wigner functionals are defined in a way equivalent to the bosonic case. As an application, we consider a two-level fermion system.
\end{abstract}

\submitto{\JPA}

\section{\label{intro}Introduction}

The tools for non-relativistic quantum physics, which is used to model, analyze, and design quantum information systems \cite{nc} in terms of quantum optics \cite{mw}, allow the description of exotic quantum states subjected to exotic quantum measurements. Such tools often use phase space representations of states and operators, expressed in terms of ladder operators or quadrature operators \cite{contvar1,weedbrook,contvar2}. Such representations are thus inspired by the Moyal formalism \cite{groenewold,moyal,psqm}. However, the authentic Moyal formalism does not involve operators.

The phase space representations of states and operators are not unique. They can either be represented by Wigner functions \cite{wigner}, Glauber-Sudarshan $P$-functions \cite{husimi} or Husimi $Q$-functions \cite{sudarshan,glauber}, among others. Wigner functions were originally defined in terms of quadrature bases, while the other two are based on coherent states. All these phase space representations are related via their characteristic functions. Here, we focus on those cases where the phase space variables represent the particle-number degrees of freedom, so that they elucidate the quantum properties of the states, as opposed to implementations where the phase space variables represent the spatiotemporal degrees of freedom \cite{bastiaans1}, or the spin degrees of freedom \cite{varilly,brif}.

Recently, the Moyal formalism has been generalized to obtain a tool with which states and operations are represented as distributions on a {\em functional} phase space. The development of a functional Moyal formalism for boson fields has been quite successfully based on a generalization of quantum optics \cite{mrowc,stquad,berra}. The functional Moyal formalism for boson fields can be based on any of the representations. The benefit of the functional formulation is that it does not impose any restrictions on the spatiotemporal degrees of freedom. Such restrictions are intrinsic in those cases where operators are combined with phase space formulations, because they represent the spatiotemporal degrees of freedom in terms of a finite number of discrete modes \cite{contvar1,weedbrook,contvar2}.

In the case of fermion fields, the original work by Cahill and Glauber \cite{caglaub} used the characteristic function to define the three different Grassmann representations associated with the Husimi $Q$-functions, Glauber-Sudarshan $P$-functions, and Wigner functions, respectively. Various other fermion phase space formulations followed. The majority of them are related to the Glauber-Sudarshan $P$-functions \cite{fermipfunc,fermibarnett,fermibarnett2,polyakov}, or the Husimi $Q$-functions \cite{fermiqfunc}, all represented in terms of fermion coherent states of various kinds. A fermion phase space formulation for Wigner functions has been presented \cite{fermiwig}, based on the assumed eigenstates of the fermion field operators. Other fermion phase space formulations based on the Wigner function follow the approach of Cahill and Glauber via the characteristic function.

An approach to define fermion Wigner functions in terms of quadrature bases, as originally used for the definition of the Wigner functions for boson fields, is lacking. The reason can be found in the fact that the Majorana operators, which are the fermion analogues of the boson quadrature operators, do not share the properties of their bosonic equivalents. They anticommute with each other and do not anticommute with themselves; the boson quadrature operators do not commute with each other, but commute with themselves. The opposite behavior of the fermion equivalent leads to complications for functions of such fermion operators.

Here, we develop a functional Moyal formalism for fermion fields, using a quadrature based approach for Wigner functionals. Instead of the Majorana operators, as fermion analogues of the boson quadrature operators, we use a special set of operators to act as fermion quadrature operators. They do not anticommute with each other and anticommute with themselves, thus behaving more closely to the boson quadrature operators, enabling the development of a fermionic Wigner functional formalism. The difference of these special fermion quadrature operators is a relative spin transformation between the ladder operators in terms of which they are formed.

The fact that such operators are not Hermitian may seem to indicate that they are not valid candidates for quadrature operators. However, the inner products among their eigenstates reveal an interesting twist. The results are surprising in that the dual space of these states do not follow from the usual Hermitian adjoint. Instead, it requires a spin transformation in addition to the Hermitian adjoint. We call it the {\em fermionic adjoint}. It is found that the special set of fermion quadrature operators are selfadjoint with respect to the fermionic adjoint. As a result, the eigenvalues of these fermion quadrature operators are not real valued. Instead, they are {\em fermionic selfconjugate}. Their eigenstates thus represent fermion quadrature bases that are analogues to their bosonic equivalents.

In terms of the Hermitian adjoint, the fermion quadrature operators that we consider here behave like fermion Bogoliubov operators. Indeed, we benefit from this fact by using a fermion Bogoliubov operator approach to compute the eigenstates of the fermion quadrature operators. It facilitates the calculation of the inner products among these eigenstates. There have been several studies of fermion Bogoliubov operators \cite{araki,ruijsenaars1,ruijsenaars2,svozil}. Some of these studies are general enough to include the case of relative spin transformations between the ladder operators. However, as far as we know, none of them revealed the above mentioned fermionic dual that includes spin transformations or its significance in the formulation of a functional Grassmann phase space.

Previous fermion phase space formulations are often done for finite dimensional systems, leading to a phase space composed of a finite number of two-dimensional phase spaces, each representing a one-dimensional system. An exception is the functional Moyal formalism for fermion fields \cite{fermiwig}, which is based on assumed eigenstates of the field operators. Here, we present an infinite dimensional functional Moyal formulation in terms of Wigner functionals, defined in terms of quadrature bases. We show the full development and compute the eigenstates for the pertinent operators explicitly.

While the formulation of the fermionic Wigner functional in terms of quadrature bases is the main result, we demonstrate its use in an application. For this purpose, we solve the fermion state in a two-level fermion system, where a complex scalar field mediates transitions between two species of fermions. These two species can represent the ground state and an excited state of fermion atoms or ions, which can be modeled by a Jaynes-Cummings Hamiltonian \cite{nc}. We are interested in the evolution of the fermion state and include all the spatiotemporal degrees of freedom. Therefore, the Jaynes-Cummings Hamiltonian is generalized for this purpose. The system is treated in a semiclassical way by representing the scalar field as a classical field. We derive an evolution equation for the fermion state and show how the solution can be represented by a transformation of an arbitrary initial state.

\section{Overview}

In what follows, we'll arrive at the definitions of fermion quadrature operators given by
\begin{equation}
\eqalign{
\hat{q}_s(\mathbf{k}) & = \frac{1}{\sqrt{2}} \left[\hat{a}_s(\mathbf{k})+\hat{a}_r^{\dag}(\mathbf{k})\varepsilon_{r,s}\right] , \\
\hat{p}_s(\mathbf{k}) & = \frac{-i}{\sqrt{2}} \left[\hat{a}_s(\mathbf{k})-\hat{a}_r^{\dag}(\mathbf{k})\varepsilon_{r,s}\right] , }
\label{ferquad}
\end{equation}
in tensor notation, using the summation convention, where $\hat{a}_s(\mathbf{k})$ and $\hat{a}_s^{\dag}(\mathbf{k})$ are the fermion annihilation and creation (ladder) operators, and $\varepsilon$ is an antisymmetric spin matrix, which implements a relative spin transformation between the two ladder operators. The properties of the spin matrix can be summarized by
\begin{equation}
\varepsilon_{r,s}^*=\varepsilon_{s,r}=-\varepsilon_{r,s} , ~~~
\varepsilon_{r,t}\varepsilon_{t,s}=\delta_{r,s} .
\label{epsprop}
\end{equation}
For two-dimensional spin degrees of freedom, $\varepsilon$ is given by the Pauli $y$-matrix
\begin{equation}
\sigma_y = \left(\begin{array}{cc} 0 & -i\\ i & 0 \end{array}\right) .
\label{defeps}
\end{equation}
The fermion ladder operators are linear combinations of the quadrature operators,
\begin{equation}
\eqalign{
\hat{a}_s(\mathbf{k}) & = \frac{1}{\sqrt{2}} \left[\hat{q}_s(\mathbf{k})+ i\hat{p}_s(\mathbf{k})\right] , \\
\hat{a}_s^{\dag}(\mathbf{k}) & = \frac{1}{\sqrt{2}} \left[\hat{q}_r(\mathbf{k})- i\hat{p}_r(\mathbf{k})\right] \varepsilon_{r,s} , }
\label{ainqp}
\end{equation}
which obey Lorentz covariant anticommutation relations
\begin{equation}
\eqalign{
\{\hat{a}_s(\mathbf{k}),\hat{a}_r(\mathbf{k}')\} & = \{\hat{a}_s^{\dag}(\mathbf{k}),\hat{a}_r^{\dag}(\mathbf{k}')\} = 0 , \\
\{\hat{a}_s(\mathbf{k}),\hat{a}_r^{\dag}(\mathbf{k}')\} & = (2\pi)^3 E_{\mathbf{k}} \delta_{s,r} \delta(\mathbf{k}-\mathbf{k}') , }
\label{fermiskepabs}
\end{equation}
with
\begin{equation}
E_{\mathbf{k}} = \sqrt{m^2+|\mathbf{k}|^2} .
\label{defek}
\end{equation}
Note that $\mathbf{k}$ is the wavevector (as opposed to the momentum). Therefore, the ``energy'' $E_{\mathbf{k}}$ and the ``mass'' $m$ both have the units of inverse distance. We exclude the antifermions from this analysis. Since their ladder operators anticommute with the fermion ladder operators, they do not affect the analysis here. An equivalent formulation exists for antifermions.

The Hermitian conjugates of the operators in (\ref{ferquad}) are
\begin{equation}
\eqalign{
\hat{q}_s^{\dag}(\mathbf{k})
& = \frac{1}{\sqrt{2}} \left[\hat{a}_s^{\dag}(\mathbf{k})-\hat{a}_r(\mathbf{k})\varepsilon_{r,s}\right]
 = -i\hat{p}_r(\mathbf{k})\varepsilon_{r,s} , \\
\hat{p}_s^{\dag}(\mathbf{k})
& = \frac{i}{\sqrt{2}} \left[\hat{a}_s^{\dag}(\mathbf{k})+\hat{a}_r(\mathbf{k})\varepsilon_{r,s}\right]
 = i\hat{q}_r(\mathbf{k})\varepsilon_{r,s} . }
\label{adferquad}
\end{equation}
As shown, the Hermitian adjoints are directly related to the original pair of operators via spin transformations.

If the Hermitian adjoint is combined with a spin transformation, so that the adjoint process is defined as
\begin{equation}
\hat{a}_s(\mathbf{k}) \rightarrow \hat{a}_r^{\dag}(\mathbf{k}) \varepsilon_{r,s} , ~~~~~
\hat{a}_s^{\dag}(\mathbf{k}) \rightarrow \varepsilon_{s,r} \hat{a}_r(\mathbf{k}) ,
\label{feradj}
\end{equation}
it reproduces the same fermion quadrature operator on which it is applied. The Hermitian adjoint combined with the spin transformations is henceforth referred to as the {\em fermionic adjoint}. It follows that the fermion quadrature operators in (\ref{ferquad}) are {\em fermionic selfadjoint}, albeit not Hermitian self adjoint.

The anticommutation relations among the fermion quadrature operators and their Hermitian adjoints are
\begin{equation}
\eqalign{
\left\{\hat{q}_r(\mathbf{k}), \hat{q}_s(\mathbf{k})\right\} & =
\left\{\hat{q}_r^{\dag}(\mathbf{k}), \hat{q}_s^{\dag}(\mathbf{k})\right\} = 0 , \\
\left\{\hat{p}_r(\mathbf{k}), \hat{p}_s(\mathbf{k})\right\} & =
\left\{\hat{p}_r^{\dag}(\mathbf{k}), \hat{p}_s^{\dag}(\mathbf{k})\right\} = 0 , \\
\left\{\hat{q}_r(\mathbf{k}), \hat{p}_s^{\dag}(\mathbf{k})\right\} & =
\left\{\hat{p}_r(\mathbf{k}), \hat{q}_s^{\dag}(\mathbf{k})\right\} = 0 , \\
\left\{\hat{q}_r(\mathbf{k}), \hat{q}_s^{\dag}(\mathbf{k})\right\} & =
\left\{\hat{p}_r(\mathbf{k}), \hat{p}_s^{\dag}(\mathbf{k})\right\}
= (2\pi)^3 E_{\mathbf{k}} \delta_{r,s} \delta(\mathbf{k}-\mathbf{k}') = \mathbf{1} , \\
\left\{\hat{q}_r(\mathbf{k}), \hat{p}_s(\mathbf{k})\right\} & =
\left\{\hat{q}_r^{\dag}(\mathbf{k}), \hat{p}_s^{\dag}(\mathbf{k})\right\}
= i (2\pi)^3 E_{\mathbf{k}} \varepsilon_{r,s} \delta(\mathbf{k}-\mathbf{k}') = i\mathbf{1} . }
\label{fermicomms}
\end{equation}

\section{Fermion Bogoliubov operators}

The anticommutation relation between each fermion quadrature operator and its Hermitian adjoint indicates that they are a pair of Bogoliubov operators. So, these fermion quadrature operators are elements of a special set of Bogoliubov operators, composed of ladder operators with a relative spin transformation. They are given by
\begin{equation}
\eqalign{
\hat{b}_s(\mathbf{k}) & = u\hat{a}_s(\mathbf{k})+v\hat{a}_r^{\dag}(\mathbf{k})\varepsilon_{r,s} , \\
\hat{b}_s^{\dag}(\mathbf{k}) & = u^* \hat{a}_s^{\dag}(\mathbf{k}) - v^*\hat{a}_r(\mathbf{k})\varepsilon_{r,s} , }
\label{ferbog}
\end{equation}
where $u$ and $v$ are complex c-numbers, related by $|u|^2+|v|^2=1$. These Bogoliubov operators have the same anticommutation relations as given in (\ref{fermiskepabs}) for the fermion ladder operators. We'll use these Bogoliubov operators to compute the eigenstates of the fermion quadrature operators.

\subsection{Eigenstates}
\label{eietoes}

The eigenstates of the fermion Bogoliubov operators are {\em displaced squeezed vacuum states}, that can be represent as
\begin{equation}
\ket{\alpha;\xi} = \hat{D}(\alpha) \hat{S}(\xi) \ket{{\rm vac}} , ~~~~~
\bra{\alpha;\xi} = \bra{{\rm vac}} \hat{S}^{\dag}(\xi) \hat{D}^{\dag}(\alpha) ,
\end{equation}
in terms of the fermion displacement operator $\hat{D}(\alpha)$ and a fermion squeezing operator $\hat{S}(\xi)$. The eigenvalue equations for the right- and left-eigenstates of a Bogoliubov operator are then given by
\begin{equation}
\hat{b}_s(\mathbf{k}) \ket{\alpha;\xi} = \ket{\alpha;\xi} b_s(\mathbf{k}) , ~~~~~
\bra{\alpha;\bar{\xi}} \hat{b}_s(\mathbf{k}) = b_s(\mathbf{k}) \bra{\alpha;\bar{\xi}} .
\label{eievgls}
\end{equation}
Here, $\xi$ and $\bar{\xi}$ are different squeezing parameters, indicating that the right- and left-eigenstates are not in general Hermitian duals of each other. Although these eigenstates incorporate the spatiotemporal and spin degrees of freedom, they do not explicitly depend on the spatiotemporal and spin degrees of freedom. Due to the anticommuting property of the fermion ladder operators, the eigenvalue function $b_s(\mathbf{k})$ must also have an anticommuting property. Hence, it is a {\em Grassmann function}.

The definition of the fermion displacement operator is \cite{caglaub}
\begin{equation}
\hat{D}(\alpha) = \exp\left(\hat{a}^{\dag}\diamond\alpha-\alpha^*\diamond\hat{a}\right) ,
\label{verp}
\end{equation}
where the $\diamond$-contractions mean
\begin{equation}
\hat{a}^{\dag}\diamond\alpha = \int \hat{a}_s^{\dag}(\mathbf{k})
\alpha_s(\mathbf{k})\ \frac{{\rm d}^3 k}{(2\pi)^3 E_{\mathbf{k}}} , ~~~~~
\alpha^*\diamond\hat{a} = \int \alpha_s^*(\mathbf{k}) \hat{a}_s(\mathbf{k})\ \frac{{\rm d}^3 k}{(2\pi)^3 E_{\mathbf{k}}} ,
\end{equation}
with $\alpha_s(\mathbf{k})$ and $\alpha_s^*(\mathbf{k})$ being Grassmann functions. The coherent states produced by this displacement operator $\ket{\alpha}=\hat{D}(\alpha) \ket{{\rm vac}}$ are equivalent to those defined in \cite{caglaub}.

The fermion squeezing operator is defined as
\begin{equation}
\hat{S}(\xi) = \exp\left(\xi\hat{R}^{\dag}-\xi^*\hat{R}\right) ,
\label{druk}
\end{equation}
where the squeezing parameter $\xi$ is a complex c-number and
\begin{equation}
\eqalign{
\hat{R} & = \case{1}{2} \int \hat{a}_s(\mathbf{k}) \varepsilon_{s,r} \hat{a}_r(\mathbf{k})\ \frac{{\rm d}^3 k}{(2\pi)^3 E_{\mathbf{k}}}
= \case{1}{2}\hat{a}\fdia\hat{a} , \\
\hat{R}^{\dag} & = \case{1}{2} \int \hat{a}_s^{\dag}(\mathbf{k}) \varepsilon_{s,r}
\hat{a}_r^{\dag}(\mathbf{k})\ \frac{{\rm d}^3 k}{(2\pi)^3 E_{\mathbf{k}}}
= \case{1}{2}\hat{a}^{\dag}\fdia\hat{a}^{\dag} . }
\label{rrc}
\end{equation}
Here, we define a $\fdia$-contraction that incorporates the spin transformation matrix $\varepsilon$ defined in (\ref{epsprop}). The latter provides the antisymmetry that allows us to define operators as products of two fermion creation operators or two fermion annihilation operators.

Contractions among pairs of fermion ladder operators and/or Grassmann functions produce {\em Grassmann even} quantities, obeying commutation relations. The full algebra of these quantities is provided in \ref{boscomm}.

The fermion displacement operator in (\ref{verp}) and the fermion squeezing operator in (\ref{druk}) are unitary under the Hermitian adjoint. Therefore, we can express the eigenvalue equation for the right-eigenstate as
\begin{equation}
\hat{S}^{\dag}(\xi) \hat{D}^{\dag}(\alpha) \hat{b}_s(\mathbf{k}) \hat{D}(\alpha) \hat{S}(\xi) \ket{{\rm vac}}
= \ket{{\rm vac}} b_s(\mathbf{k}) .
\label{bogeief}
\end{equation}
The commutation relations
\begin{equation}
\eqalign{
\left[u\hat{a}_s+v\hat{a}_r^{\dag}\varepsilon_{r,s}, \hat{a}^{\dag}\diamond\alpha-\alpha^*\diamond\hat{a}\right]
= u\alpha_s+v\alpha_r^*\varepsilon_{r,s} , \\
\left[u\hat{a}_s+v\hat{a}_r^{\dag}\varepsilon_{r,s}, \xi\hat{R}^{\dag}-\xi^*\hat{R}\right]
= \xi^* v\hat{a}_s-\xi u\hat{a}_r^{\dag}\varepsilon_{r,s} , }
\end{equation}
together with the identity
\begin{eqnarray}
\exp(\hat{X})\hat{Y}\exp(-\hat{X})
& = \hat{Y} + [\hat{X},\hat{Y}] + \frac{1}{2!}[\hat{X},[\hat{X},\hat{Y}]] \nonumber \\
& + \frac{1}{3!}[\hat{X},[\hat{X},[\hat{X},\hat{Y}]]] + ...\ ,
\label{eksopprod}
\end{eqnarray}
lead to
\begin{equation}
\eqalign{
\hat{D}^{\dag}(\alpha) \hat{b}_s \hat{D}(\alpha)
& = \hat{b}_s + u\alpha_s+v\alpha_r^*\varepsilon_{r,s} , \\
\hat{S}^{\dag}(\xi)\hat{b}_s\hat{S}(\xi) & = [u\cos(|\xi|)+v\Phi^*\sin(|\xi|)] \hat{a}_s \\
& \ \ + [v\cos(|\xi|)-u\Phi\sin(|\xi|)] \hat{a}_r^{\dag}\varepsilon_{r,s} , }
\end{equation}
where we represent the squeezing parameter as $\xi=|\xi|\Phi$. These expansions allow us to solve (\ref{bogeief}). We obtain the eigenvalue function in terms of $u$ and $v$ as
\begin{equation}
b_s(\mathbf{k}) = u\alpha_s(\mathbf{k})+v\alpha_r^*(\mathbf{k})\varepsilon_{r,s} ,
\label{bogeiew}
\end{equation}
with
\begin{equation}
u = \chi\cos(|\xi|) ~~~ {\rm and} ~~~ v = \chi\Phi\sin(|\xi|) ,
\end{equation}
where $\chi$ is an arbitrary complex c-number. We choose $\chi^2=\Phi^*$ and define an angle $\psi=|\xi|$ to get
\begin{equation}
u = \chi\cos(\psi) ~~~ {\rm and} ~~~ v = \chi^*\sin(\psi) .
\label{defuv}
\end{equation}
There is no restriction on the Grassmann parameter function of the displacement $\alpha_s(\mathbf{k})$.

For the case of the left-eigenstate $\langle\alpha;\bar{\xi}|$ with $\bar{\xi}=|\bar{\xi}|\bar{\Phi}$, it is required that
\begin{equation}
u\cos(|\bar{\xi}|)+v\bar{\Phi}^*\sin(|\bar{\xi}|) = 0 .
\end{equation}
Together with (\ref{defuv}), this requirement leads to two possible sets of relationships for the phase factors and angles
\begin{equation}
\bar{\Phi}^*=\pm\Phi^*=\pm\chi^2 ~~~~~ {\rm and} ~~~~~
\bar{\psi}=\pm\psi-\case{\pi}{2} ,
\label{barpsi}
\end{equation}
where we defined $|\bar{\xi}|=\bar{\psi}$. The upper signs and lower signs go together for each solution. While $\alpha_s(\mathbf{k})$ is unrestricted, the squeezing parameters of the right- and left-eigenstates are respectively given in terms of the parameters in (\ref{defuv}) by
\begin{equation}
\xi = \psi (\chi^*)^2 ~~~~~ {\rm and} ~~~~~
\bar{\xi} = (\psi\mp\case{\pi}{2}) (\chi^*)^2 .
\label{barxi}
\end{equation}

We can follow the same procedure to obtain the right- and left-eigenstates of $\hat{b}_s^{\dag}(\mathbf{k})$ and the eigenvalue function $\bar{b}_s(\mathbf{k})$, as found in the eigenvalue equations
\begin{equation}
\hat{b}_s^{\dag}(\mathbf{k}) \ket{\alpha;\bar{\xi}} = \ket{\alpha;\bar{\xi}} \bar{b}_s(\mathbf{k}) , ~~~~~
\bra{\alpha;\xi} \hat{b}_s^{\dag}(\mathbf{k}) = \bar{b}_s(\mathbf{k}) \bra{\alpha;\xi} .
\end{equation}
However, these eigenvalue equations are the Hermitian adjoints of those for $\hat{b}_s(\mathbf{k})$. Therefore, the right- and left-eigenstates of $\hat{b}_s^{\dag}(\mathbf{k})$ are the Hermitian duals of the left- and right-eigenstates of $\hat{b}_s(\mathbf{k})$, respectively (which is the reason for the assignments of $\xi$ and $\bar{\xi}$), and the Grassmann eigenvalue function $\bar{b}_s(\mathbf{k})$ is the complex conjugate of $b_s(\mathbf{k})$
\begin{equation}
\bar{b}_s(\mathbf{k}) = u^*\alpha_r^*(\mathbf{k}) - v^*\alpha_s(\mathbf{k})\varepsilon_{r,s} .
\label{bogeiewc}
\end{equation}

\subsection{Inner products}
\label{inprod}

The inner products among the different eigenstates reveal important properties. The generic form of such inner products between left- and right-eigenstates is
\begin{eqnarray}
\braket{\alpha_1,\xi_1}{\alpha_2,\xi_2} & = \bra{{\rm vac}} \exp\left(-\xi_1\hat{R}^{\dag}+\xi_1^*\hat{R}\right)
\exp\left(-\hat{\alpha}_1^{\dag}+\hat{\alpha}_1\right) \nonumber \\
& \times \exp\left(\hat{\alpha}_2^{\dag}-\hat{\alpha}_2\right)
\exp\left(\xi_2\hat{R}^{\dag} - \xi_2^*\hat{R}\right) \ket{{\rm vac}} \nonumber \\
& = \Psi \bra{{\rm vac}} \exp\left(-\xi_1\hat{R}^{\dag}+\xi_1^*\hat{R}\right)
\exp\left(\hat{\eta}^{\dag}-\hat{\eta}\right) \nonumber \\
& \times \exp\left(\xi_2\hat{R}^{\dag} - \xi_2^*\hat{R}\right) \ket{{\rm vac}} ,
\label{oorvds}
\end{eqnarray}
where $\hat{\alpha}_n=\alpha_n^*\diamond\hat{a}$, $\hat{\alpha}_n^{\dag}=\hat{a}^{\dag}\diamond\alpha_n$, and
\begin{equation}
\eqalign{
\Psi & = \exp\left(\case{1}{2} \alpha_1^*\diamond\alpha_2 -\case{1}{2} \alpha_2^*\diamond\alpha_1\right) , \\
\hat{\eta} & = \eta^*\diamond\hat{a} = (\alpha_2^*-\alpha_1^*)\diamond\hat{a} , \\
\hat{\eta}^{\dag} & = \hat{a}^{\dag}\diamond\eta = \hat{a}^{\dag}\diamond(\alpha_2-\alpha_1) . }
\end{equation}

In order to compute the generic inner product, we need to convert (\ref{oorvds}) to normal order. As a result, a more complex expression is obtained with new operators generated by the current ones. To aid the commutation process, we introduce an auxiliary variable $t$. It then reads
\begin{eqnarray}
& \exp\left(-t\xi_1\hat{R}^{\dag}+t\xi_1^*\hat{R}\right)
\exp\left(t\hat{\eta}^{\dag}-t\hat{\eta}\right)
\exp\left(t\xi_2\hat{R}^{\dag} - t\xi_2^*\hat{R}\right) \nonumber \\
& = \exp[h_0(t)] \exp[h_1(t)\hat{\eta}^{\dag}] \exp[h_2(t)\hat{\eta}_{\varepsilon}^{\dag}] \exp[h_3(t)\hat{R}^{\dag}] \nonumber \\
& \times \exp[h_4(t)\hat{s}] \exp[h_5(t)\hat{\eta}_{\varepsilon}] \exp[h_6(t)\hat{\eta}] \exp[h_7(t)\hat{R}] ,
\label{omksdt}
\end{eqnarray}
where $\hat{\eta}_{\varepsilon}=\eta\fdia\hat{a}$ and $\hat{\eta}_{\varepsilon}^{\dag}=\hat{a}^{\dag}\fdia\eta^*$, with $\fdia$ defined in (\ref{rrc}), and $h_n(t)$ represents unknown functions that go to zero for $t=0$. We only have one Grassmann parameter function and its complex conjugate on the left-hand side. Therefore, we expect all the Grassmann parameter functions on the right-hand side to be related to them.

Substituting (\ref{omksdt}) into (\ref{oorvds}), we get
\begin{equation}
\braket{\alpha_1,\xi_1}{\alpha_2,\xi_2} = \Psi \exp[h_0(t)]\exp\left[-\case{1}{2}h_4(t)\Omega\right] ,
\label{oorvds1}
\end{equation}
where we used (\ref{numr}). Hence, only two $h$-functions are relevant for the final result.

Next, we apply a derivative with respect to $t$ on both sides of (\ref{omksdt}) and then remove as many of the exponentiated operators by operating on both sides of the equations with the respective inverse operators applied on their right-hand sides. The result is simplified with the aid of the identities in \ref{boseksop}. It separates into a set of eight differential equations, associated with the different operators. These differential equations can be further simplified and solved to produce solutions for all the $h$-functions.

The solutions for $h_4(t)$ and $h_0(t)$ are given by
\begin{equation}
\eqalign{
h_0(t) & = \case{1}{2}t^2 \left\{ \left[\Phi_1^*\Phi_2\sin(t|\xi_1|)\sin(t|\xi_2|)-\cos(t|\xi_1|)\cos(t|\xi_2|)\right]
\eta^*\diamond\eta \right. \\
& \left. +\Phi_2\cos(t|\xi_1|)\sin(t|\xi_2|) \eta^*\fdia\eta^*+\Phi_1^*\sin(t|\xi_1|)\cos(t|\xi_2|) \eta\fdia\eta \right\} \\
& \times \left[\cos(t|\xi_1|)\cos(t|\xi_2|)+\Phi_1^*\Phi_2\sin(t|\xi_1|)\sin(t|\xi_2|)\right]^{-1} , \\
h_4(t) & = -\ln\left[\cos(t|\xi_1|)\cos(t|\xi_2|)+\Phi_1^*\Phi_2\sin(t|\xi_1|)\sin(t|\xi_2|)\right] . }
\label{opls}
\end{equation}
where $\xi_x=|\xi_x|\Phi_x$ for $x=1,2$. Substituted into (\ref{oorvds1}) with $t=1$, they produce the expression for the generic inner product, which can be used to compute the inner product between any two eigenstates. It reads
\begin{eqnarray}
\fl \braket{\alpha_1,\xi_1}{\alpha_2,\xi_2} & = (C_1C_2+S_1S_2)^{\Omega/2} \Psi \nonumber \\
& \times \exp\left[\frac{(S_1S_2 - C_1C_2) \eta^*\diamond\eta
+ C_1S_2 \eta^*\fdia\eta^* + S_1C_2 \eta\fdia\eta}{2(C_1C_2+S_1S_2)}\right] \label{oorvfin1} \\
& = (C_1C_2+S_1S_2)^{\Omega/2}
\Psi \exp\left[\frac{(C_2\eta+S_2 \eta^*\cdot\varepsilon)\fdia(S_1\eta+C_1\varepsilon\cdot\eta^*)}
{2(C_1C_2+S_1S_2)}\right] , \label{oorvfin2}
\end{eqnarray}
where we simplified the notation for the spin sums in (\ref{oorvfin2}) to represent $\varepsilon_{s,r}\eta_r^*$ and $\eta_r^*\varepsilon_{r,s}$ as $\varepsilon\cdot\eta^*$ and $\eta^*\cdot\varepsilon$, respectively, so that we can use the compact notation involving $\fdia$. We also defined
\begin{equation}
\eqalign{
C_1 & = \cos(|\xi_1|) , ~~~ S_1 = \Phi_1^*\sin(|\xi_1|) , \\
C_2 & = \cos(|\xi_2|) , ~~~ S_2 = \Phi_2\sin(|\xi_2|) . }
\label{defcspar}
\end{equation}

Note that, if the displacements of the two states are the same, then $\eta=\alpha_2-\alpha_1=0$ and $\Psi=1$, so that
\begin{equation}
\braket{\alpha_1,\xi_1}{\alpha_1,\xi_2} = (C_1C_2+S_1S_2)^{\Omega/2} .
\end{equation}
If we also have $\xi_2=\xi_1$, then $C_2=C_1$ and $S_2=S_1^*$, so that
\begin{equation}
\braket{\alpha_1,\xi_1}{\alpha_1,\xi_1} = 1 .
\end{equation}
Hence, all the states are normalized as expected, provided that the dual is defined with the Hermitian adjoint.

An important case is where $\Phi_1=\Phi_2=1$. Then
\begin{equation}
C_1 C_2 \pm S_1 S_2 = \cos(|\xi_1| \mp |\xi_2|) .
\end{equation}
When $|\xi_1|-|\xi_2|=\pm\case{\pi}{2}$, as when the inner product involves the left- and right-eigenstates of the same operator, we get $C_1 C_2 + S_1 S_2=0$. It produces a singularity, which requires the limit process considered in section~\ref{binprod}. More importantly, if we then also have $|\xi_1|+|\xi_2|=0$, so that $C_1 C_2 - S_1 S_2=1$, the limit would produce a product of differences between the respective Grassmann parameter functions that represents a Grassmann Dirac delta functional. The opposite situation where $C_1 C_2 + S_1 S_2=1$ and $C_1 C_2 - S_1 S_2=0$, which happens when the inner product involves the right-eigenstates and their Hermitian duals with different displacements, produces a result that behaves like a phase factor; its Hermitian conjugate changes sign in the exponent.

These results are counterintuitive: the Grassmann Dirac delta functional represents an orthogonality condition, which is expected to exist between right-eigenstates and their Hermitian duals, and the phase factors indicate mutually unbiased bases, which are not expected for eigenstates and their Hermitian duals. This unexpected behavior justifies the notion of a fermionic dual, which redefines the inner product. It is explained in more detail in section~\ref{fqo}.

\subsection{Fermionic dual eigenstates}

The eigenstates computed in section~\ref{eietoes} are obtained under the assumption that the dual space is given by the Hermitian adjoint without spin transformations. If the dual space is given by the fermionic adjoint instead, the Bogoliubov operators and their eigenstates would in general be different. The reason is that unitary operators under the Hermitian adjoint are not necessarily unitary under the fermionic adjoint. The squeezing operators are defined differently under the fermionic adjoint. (When the fermionic adjoint is extended to the Grassmann parameter functions, the displacement operators would transform the same under both adjoints).

Eigenstates that are defined with squeezing operators that are unitary under the fermionic adjoint, produce relationships between the Bogoliubov coefficients $u$ and $v$ and the squeezing parameter $\xi$ that are more similar to the bosonic case: they are parameterized as hyperbolic trigonometric functions. It implies that one Bogoliubov coefficient is always bigger than the other. Such Bogoliubov operators only have either right-eigenstates or left-eigenstates satisfying (\ref{eievgls}), depending on which coefficient is bigger, but not both. (Note that these eigenvalue equations assume eigenstates that do not explicitly depend on the spin and wavevector. If such dependencies are allowed, both right- and left-eigenstates can be obtained, but we are not interested in such cases.)

Since these operators only have one-sided eigenstates, they cannot be fermionic selfadjoint. Therefore, they are not candidates for fermion quadrature operators. As a result, we won't consider them any further.

\section{Fermion quadrature operators}
\label{fqo}

\subsection{Operator definitions}

We are looking for fermion quadrature operators that are equivalent to the quadrature operators of boson fields. Such operators would have coefficients for the two terms with the same magnitude. Under the Hermitian adjoint, we obtained Bogoliubov operators having both left- and right-eigenstates. While these operators are not Hermitian self adjoint, we find that they can have coefficients with the same magnitude and be fermionic selfadjoint for certain parameter values.

The Bogoliubov operator in (\ref{ferbog}), with its coefficients given by (\ref{defuv}), becomes
\begin{equation}
\hat{b}_s(\mathbf{k}) = \chi\cos(\psi) \hat{a}_s(\mathbf{k}) + \chi^*\sin(\psi) \hat{a}_r^{\dag}(\mathbf{k}) \varepsilon_{r,s} .
\label{defhb}
\end{equation}
When we apply to this operator the fermionic adjoint, defined in (\ref{feradj}), it becomes
\begin{eqnarray}
\hat{b}_s^{\ddag}(\mathbf{k}) & = \chi^*\cos(\psi) \hat{a}_r^{\dag}(\mathbf{k}) \varepsilon_{r,s}
+ \chi \sin(\psi) \varepsilon_{r,t} \hat{a}_t(\mathbf{k}) \varepsilon_{r,s}^* \nonumber \\
& = \chi\sin(\psi) \hat{a}_s(\mathbf{k}) + \chi^*\cos(\psi) \hat{a}_r^{\dag}(\mathbf{k}) \varepsilon_{r,s} ,
\end{eqnarray}
where $\ddag$ represents the fermionic adjoint. This operator is fermionic self adjoint for $\cos(\psi)=\sin(\psi)$, implying $\psi=\case{\pi}{4}$.

\subsection{Eigenstates}

Another requirement for the Bogoliubov operator to be fermionic selfadjoint is that its left-eigenstate
\begin{equation}
\bra{\alpha;\bar{\xi}} = \bra{{\rm vac}} \exp\left(-\bar{\xi}\hat{R}^{\dag}+\bar{\xi}^*\hat{R}\right) \hat{D}^{\dag}(\alpha) ,
\label{regc}
\end{equation}
must be the fermionic adjoint of its right-eigenstate
\begin{equation}
\ket{\alpha;\xi} = \hat{D}(\alpha) \exp\left(\xi\hat{R}^{\dag}-\xi^*\hat{R}\right) \ket{{\rm vac}} .
\label{reg}
\end{equation}
It is only possible if the fermionic adjoint is extended to the Grassmann parameter functions, so that the displacement operators would transform the same in both cases. To define a {\em fermionic conjugation} for them, we separate the Grassmann functions and their complex conjugates into two disjoint sets $\mathcal{G}$ and $\mathcal{G}^*$. If $\alpha\in\mathcal{G}$, then $\alpha^*\in\mathcal{G}^*$. (A similar treatment is provided in \cite{fermibarnett}. There, the complex conjugates $g^*$ are replaced by $g^+$. Here, we retain the traditional complex conjugate notation.) The fermionic conjugation process for the Grassmann functions is then defined by the following mappings between the elements of these sets
\begin{equation}
\alpha_s(\mathbf{k}) \rightarrow \alpha_r^*(\mathbf{k}) \varepsilon_{r,s} \equiv \alpha^*\cdot\varepsilon , ~~~~~
\alpha_s^*(\mathbf{k}) \rightarrow \varepsilon_{s,r} \alpha_r(\mathbf{k}) \equiv \varepsilon\cdot\alpha .
\label{ferconj}
\end{equation}
The fermionic conjugation and the fermionic adjoint lead to the transformations of the different contractions provided in \ref{herfer}.
It follows that
\begin{equation}
\hat{D}^{\ddag}(\alpha) = \hat{D}^{\dag}(\alpha) = \exp\left(-\hat{a}^{\dag}\diamond\alpha+\alpha^*\diamond\hat{a}\right) .
\label{defhverp}
\end{equation}
For the squeezing operator, we have $\hat{R}^{\ddag}=-\hat{R}^{\dag}$, so that
\begin{equation}
\eqalign{
\hat{S}^{\dag}(\xi) & = \exp\left(-\xi\hat{R}^{\dag}+\xi^*\hat{R}\right) , \\
\hat{S}^{\ddag}(\xi) & = \exp\left(\xi\hat{R}^{\dag}-\xi^*\hat{R}\right) \equiv \hat{S}(\xi) . }
\end{equation}
Hence, the fermionic adjoint of the right-eigenstate would be equal to the left-eigenstate when the arguments of the respective squeezing operators are equal
\begin{equation}
-\bar{\xi}\hat{R}^{\dag}+\bar{\xi}^*\hat{R} = \xi\hat{R}^{\dag}-\xi^*\hat{R} .
\end{equation}
It means that
\begin{equation}
\bar{\xi}=\bar{\psi}\bar{\Phi}=(\psi\mp\case{\pi}{2})\Phi = -\xi=-\psi\Phi ,
\end{equation}
where we used the relationships in (\ref{barxi}). It implies that $\psi=\pm\case{\pi}{4}$, which is consistent with the previous condition for the upper sign. The phase factor $\Phi=(\chi^*)^2$ is not restricted by either of the conditions. We use two choices: $\chi=1$ and $\chi=-i$, leading to $\Phi=1$ and $\Phi=-1$, respectively. When these parameters are substituted into (\ref{defhb}), they produce the fermion quadrature operators $\hat{q}$ and $\hat{p}$ in (\ref{ferquad}).

For the eigenstates, the two choices give
\begin{equation}
\left\{\xi = \case{\pi}{4} , \bar{\xi} = -\case{\pi}{4}\right\} ~~~ {\rm and} ~~~
\left\{\xi = -\case{\pi}{4} , \bar{\xi} = \case{\pi}{4}\right\} ,
\end{equation}
for $\hat{q}$ and $\hat{p}$, respectively. The right- and left-eigenstates of $\hat{q}_s(\mathbf{k})$ are given by
\begin{equation}
\eqalign{
\ket{q} & = \ket{\alpha;\case{\pi}{4}} = \hat{D}(\alpha) \hat{S}(\case{\pi}{4}) \ket{{\rm vac}} , \\
\bra{q} & = \bra{\alpha;-\case{\pi}{4}} = \bra{{\rm vac}} \hat{S}^{\dag}(-\case{\pi}{4}) \hat{D}^{\dag}(\alpha) , }
\label{rlq}
\end{equation}
and those of $\hat{p}_s(\mathbf{k})$ are
\begin{equation}
\eqalign{
\ket{p} & = \ket{\alpha;-\case{\pi}{4}} = \hat{D}(\alpha) \hat{S}(-\case{\pi}{4}) \ket{{\rm vac}} , \\
\bra{p} & = \bra{\alpha;\case{\pi}{4}} = \bra{{\rm vac}} \hat{S}^{\dag}(\case{\pi}{4}) \hat{D}^{\dag}(\alpha) , }
\label{rlp}
\end{equation}
where
\begin{eqnarray}
\hat{S}(\case{\pi}{4}) & = \hat{S}^{\dag}(-\case{\pi}{4})
= \exp\left(\case{1}{4}\pi\hat{R}^{\dag}-\case{1}{4}\pi\hat{R}\right) , \\
\hat{S}^{\dag}(\case{\pi}{4}) & = \hat{S}(-\case{\pi}{4})
= \exp\left(-\case{1}{4}\pi\hat{R}^{\dag}+\case{1}{4}\pi\hat{R}\right) .
\end{eqnarray}

The respective eigenvalue functions are
\begin{equation}
\eqalign{
q_s(\mathbf{k}) & = \frac{1}{\sqrt{2}} \left[\alpha_s(\mathbf{k})+\alpha_r^*(\mathbf{k})\varepsilon_{r,s}\right] , \\
p_s(\mathbf{k}) & = \frac{-i}{\sqrt{2}} \left[\alpha_s(\mathbf{k})-\alpha_r^*(\mathbf{k})\varepsilon_{r,s}\right] . }
\label{eiewaar}
\end{equation}
Although these eigenvalue functions are not real-valued Grassmann functions (the fermionic conjugation always produces complex values due to the spin transformations), they are {\em fermionic selfconjugate}, which can be readily confirmed with the aid of (\ref{ferconj}). They do not transform in the same way under fermionic conjugation as the complex-valued Grassmann parameter functions because each is composed of both a complex-valued Grassmann parameter function and its complex conjugate. So, we need to treat the Grassmann parameter functions and Grassmann eigenvalue functions as distinct sets of Grassmann functions, transforming differently under fermionic conjugation.

\subsection{Inner products}
\label{binprod}

The squeezing parameters in the eigenstates have only two possible values $\xi=\pm\case{\pi}{4}$, and as such can produce singularities, as mentioned at the end of section~\ref{inprod}. To address such a singularity in the expression of the inner product, we instead use $\xi=\pm(\case{\pi}{4}-\case{1}{2}\epsilon)$, where $\epsilon$ is taken to zero in the limit. So, the parameters in (\ref{defcspar}) become
\begin{equation}
C(\pm\case{\pi}{4}) = \cos(\case{\pi}{4}-\case{1}{2}\epsilon) ~~~ {\rm and} ~~~
S(\pm\case{\pi}{4}) = \pm\sin(\case{\pi}{4}-\case{1}{2}\epsilon) .
\end{equation}
For the two selfoverlaps, we use (\ref{oorvfin2}) with
\begin{equation}
\eqalign{
(C_1C_2+S_1S_2) = \sin(\epsilon) \approx \epsilon , \\
C_1 = C_2 = \case{1}{\sqrt{2}} , ~~~
S_1 = \mp \case{1}{\sqrt{2}} , ~~~
S_2 = \pm \case{1}{\sqrt{2}} , }
\end{equation}
where the upper signs are for $\braket{q}{q'}$ and the lower signs for $\braket{p}{p'}$. Hence,
\begin{equation}
\eqalign{
\braket{q}{q'} & = \lim_{\epsilon\rightarrow 0}\
\braket{\alpha_1;-\case{\pi}{4}+\case{1}{2}\epsilon}{\alpha_2;\case{\pi}{4}-\case{1}{2}\epsilon} \\
& = \Psi \lim_{\epsilon\rightarrow 0} \epsilon^{\Omega/2}
\exp\left[\frac{-(\eta+\eta^*\cdot\varepsilon)\fdia(\eta-\varepsilon\cdot\eta^*)}{4\epsilon}\right] , \\
\braket{p}{p'} & = \lim_{\epsilon\rightarrow 0}\
\braket{\alpha_1;\case{\pi}{4}-\case{1}{2}\epsilon}{\alpha_2;-\case{\pi}{4}+\case{1}{2}\epsilon} \\
& = \Psi \lim_{\epsilon\rightarrow 0} \epsilon^{\Omega/2}
\exp\left[\frac{(\eta-\eta^*\cdot\varepsilon)\fdia(\eta+\varepsilon\cdot\eta^*)}{4\epsilon}\right] . }
\label{eieoorv}
\end{equation}
The cross-overlaps are expressed with (\ref{oorvfin1}), for which
\begin{equation}
\eqalign{
(C_1C_2+S_1S_2) = 1 , \\
(C_1C_2-S_1S_2) = \sin(\epsilon) \approx \epsilon , \\
C_1S_2 = \mp \case{1}{2} \sin(\case{\pi}{2}-\epsilon) \approx \mp \case{1}{2} , \\
S_1C_2 = \mp \case{1}{2} \sin(\case{\pi}{2}-\epsilon) \approx \mp \case{1}{2} , }
\end{equation}
where the upper signs are for $\braket{q}{p}$ and the lower signs are for $\braket{p}{q}$. So the overlaps are
\begin{equation}
\eqalign{
\braket{q}{p} & = \lim_{\epsilon\rightarrow 0}\
\braket{\alpha_1;-\case{\pi}{4}+\case{1}{2}\epsilon}{\alpha_2;-\case{\pi}{4}+\case{1}{2}\epsilon}
= \Psi \exp\left(-\case{1}{4}\eta^*\fdia\eta^*-\case{1}{4}\eta\fdia\eta\right) , \\
\braket{p}{q} & = \lim_{\epsilon\rightarrow 0}\
\braket{\alpha_1;\case{\pi}{4}-\case{1}{2}\epsilon}{\alpha_2;\case{\pi}{4}-\case{1}{2}\epsilon}
= \Psi \exp\left(\case{1}{4}\eta^*\fdia\eta^*+\case{1}{4}\eta\fdia\eta\right) . }
\label{kruisoorv}
\end{equation}
Only one term in the expansion of the exponential function in the selfoverlaps survives the limit. It is the term consisting of the product of all Grassmann numbers ($\Omega$ of them). In this term, the $\epsilon$ factors cancel and the limit can be taken without effect. All lower order terms have extra factors of $\epsilon$, becoming zero in the limit. Higher order terms (for finite $\Omega$) have higher powers of the same Grassmann numbers leading to zero.

\subsection{Field variables}

So far, we have expressed the inner products in terms of the complex Grassmann functions that parameterize the displacements of the eigenstates. In preparation for the functional phase space representations, we convert these expressions so that they are represented in terms of the Grassmann eigenvalue functions, which will act as {\em field variables} in the functionals.

When we substitute $\eta=\alpha_2-\alpha_1$ into the exponents of the self overlaps, they become
\begin{equation}
\eqalign{
-(\eta+\eta^*\cdot\varepsilon)\fdia(\eta-\varepsilon\cdot\eta^*) = -2(q_2-q_1)\fdia(q_2-q_1) , \\
(\eta-\eta^*\varepsilon)\fdia(\eta+\varepsilon\cdot\eta^*) = -2(p_2-p_1)\fdia(p_2-p_1) , }
\end{equation}
where we used the expressions for the eigenvalue functions in (\ref{eiewaar}) to define
\begin{equation}
\eqalign{
q_x = \case{1}{\sqrt{2}} (\alpha_x+\alpha_x^*\cdot\varepsilon) = \case{1}{\sqrt{2}} (\alpha_x-\varepsilon\cdot\alpha_x^*) , \\
p_x = -i \case{1}{\sqrt{2}} (\alpha_x-\alpha_x^*\cdot\varepsilon) = -i \case{1}{\sqrt{2}} (\alpha_x+\varepsilon\cdot\alpha_x^*) , }
\end{equation}
with $x=1,2$. Since the eigenvalue functions are fermionic selfconjugate --- i.e., $q^{\ddag}=q$ and $(q^*)^{\ddag}=q^*$ --- the contractions transform differently under the fermionic adjoint, as shown in \ref{herfer}.

The complex-valued Grassmann parameter functions, respectively associated with the $\ket{q}$'s and $\ket{p}$'s, can be expressed in terms of their eigenvalue functions, as
\begin{equation}
\alpha = \case{1}{\sqrt{2}} (q-q^*\cdot\varepsilon) , ~~~~~
\alpha = i \case{1}{\sqrt{2}} (p-p^*\cdot\varepsilon) .
\label{defbinqp}
\end{equation}
If the same Grassmann parameter function appears in both eigenvalue functions, it can then be extracted as
\begin{equation}
\alpha = \case{1}{\sqrt{2}} (q+i p) , ~~~~~
\alpha^* = \case{1}{\sqrt{2}} (q-i p)\cdot\varepsilon .
\label{qpnaa}
\end{equation}

We now use the expressions in (\ref{defbinqp}) to consider $\Psi$. For $\braket{q_1}{q_2}$, its argument is
\begin{equation}
\case{1}{2} \alpha_1^*\diamond\alpha_2 -\case{1}{2} \alpha_2^*\diamond\alpha_1
= \case{1}{2} (q_1^*\diamond q_2-q_2^*\diamond q_1) ,
\end{equation}
and for $\braket{p_1}{p_2}$, it is
\begin{equation}
\case{1}{2} \alpha_1^*\diamond\alpha_2 -\case{1}{2} \alpha_2^*\diamond\alpha_1
= \case{1}{2} (p_1^*\diamond p_2-p_2^*\diamond p_1) .
\end{equation}
Based on (\ref{fadjeie}), these arguments change sign under the fermionic adjoint. Therefore, $\Psi$ acts as a phase factor.

These results are now used to express the inner products in (\ref{eieoorv}) in terms of eigenvalue functions:
\begin{equation}
\eqalign{
\braket{q}{q'} & = \exp\left[\case{1}{2} (q_1^*\diamond q_2-q_2^*\diamond q_1)\right]
\lim_{\epsilon\rightarrow 0} \epsilon^{\Omega/2} \exp\left[\frac{-(q_2-q_1)\fdia(q_2-q_1)}{2\epsilon}\right] \\
& = \exp\left[\case{1}{2} (q_1^*\diamond q_2-q_2^*\diamond q_1)\right] \delta[q_2-q_1] , \\
\braket{p}{p'} & = \exp\left[\case{1}{2} (p_1^*\diamond p_2-p_2^*\diamond p_1)\right]
\lim_{\epsilon\rightarrow 0} \epsilon^{\Omega/2}\exp\left[\frac{-(p_2-p_1)\fdia(p_2-p_1)}{2\epsilon}\right] \\
& = \exp\left[\case{1}{2} (p_1^*\diamond p_2-p_2^*\diamond p_1)\right] \delta[p_2-p_1] , }
\label{defbineie}
\end{equation}
where we represent the limits as Grassmann Dirac delta functionals in the end. The phase factors in front can be discarded, because the Dirac delta functionals will enforce equality of the two parameter functions.

Now we consider the cross-overlaps in (\ref{kruisoorv}) and evaluate their exponents where the Grassmann parameter functions in $\eta$ are defined in terms of the eigenvalue functions, as given in (\ref{defbinqp}). The argument becomes
\begin{eqnarray}
\case{1}{4}\eta^*\fdia\eta^*+\case{1}{4}\eta\fdia\eta
& = -\case{1}{2} \left(q^*\diamond q - i q^*\fdia p^* + i q\fdia p - p^*\diamond p \right) \nonumber \\
& = -\case{1}{2}\left(q^*+i p\cdot\varepsilon\right)\diamond\left(q-i \varepsilon\cdot p^*\right) .
\end{eqnarray}
The argument of $\Psi$ for $\braket{q}{p}$ and $\braket{p}{q}$ are
\begin{equation}
\eqalign{
\case{1}{2} \alpha^*\diamond\beta -\case{1}{2} \beta^*\diamond\alpha & = i \case{1}{2} (q^*\fdia p^*+q\fdia p) , \\
\case{1}{2} \alpha^*\diamond\beta -\case{1}{2} \beta^*\diamond\alpha & = -i \case{1}{2} (q^*\fdia p^*+q\fdia p) , }
\end{equation}
respectively. These arguments are the same, apart from an overall change in sign. So, the expressions for the cross-overlaps become
\begin{equation}
\eqalign{
\braket{q}{p} & = \exp\left(i q\fdia p + \case{1}{2} q^*\diamond q - \case{1}{2} p^*\diamond p\right) , \\
\braket{p}{q} & = \exp\left(-i q\fdia p - \case{1}{2} q^*\diamond q + \case{1}{2} p^*\diamond p\right) . }
\end{equation}
All the terms in the exponents transform as phases under the fermionic adjoint. Since the two terms $\case{1}{2} q^*\diamond q$ and $\case{1}{2} p^*\diamond p$ only depend on the respective parameter functions of the two states that are being overlapped, and a global phase factor does not affect a state, we can discard these phase factors. Alternatively, we can redefine the states by absorbing these phase factors. They would then emerge in the selfoverlaps and combine with those phase factors, which would still cancel as the eigenvalue functions are set equal by the Dirac delta functional. Either way, we discard all the phase factors and represent the inner products among the different states as
\begin{equation}
\eqalign{
\braket{q}{q'} & = \delta[q-q'] , ~~~~~
\braket{q}{p} = \exp\left(i q\fdia p\right) , \\
\braket{p}{p'} & = \delta[p-p'] , ~~~~~
\braket{p}{q} = \exp\left(-i q\fdia p\right) , }
\label{binneqp}
\end{equation}
where $q$ and $p$ act as field variables.

\subsection{Completeness}

Based on the form of the orthogonality conditions in (\ref{binneqp}), the expected form of the associated completeness conditions are
\begin{equation}
\int \ket{q}\bra{q}\ \mathcal{D}[q] = \mathds{I} , ~~~~~
\int \ket{p}\bra{p}\ \mathcal{D}[p] = \mathds{I} ,
\label{volqp}
\end{equation}
where $\mathds{I}$ is the identity operator for the Hilbert space. The eigenvalue functions are now employed as Grassmann
integration field variables. These definitions allow the required idempotency of the identity operator, thanks to the orthogonality conditions.

To demonstrate the completeness, we can overlap the different resolutions of the identity operator by two arbitrary states and see whether the result is the inner product between these states. There are two ways to resolve the identity and there are four inner products in (\ref{binneqp}), leading to eight expressions. Six of them (three for each resolution of the identity operator) are easily evaluated thanks to the Grassmann Dirac delta functionals. The remaining two are
\begin{equation}
\eqalign{
\bra{p_1}\mathds{I}\ket{p_2}
& = \int \braket{p_1}{q}\braket{q}{p_2}\ \mathcal{D}[q] \\
& = \int \exp\left(-i q\fdia p_1\right) \exp\left(i q\fdia p_2\right)\ \mathcal{D}[q]
= \delta[p_1-p_2] , \\
\bra{q_1}\mathds{I}\ket{q_2}
& = \int \braket{q_1}{p}\braket{p}{q_2}\ \mathcal{D}[p] \\
& = \int \exp\left(i q_1\fdia p\right) \exp\left(-i q_2\fdia p\right)\ \mathcal{D}[p]
= \delta[q_1-q_2] . }
\label{diracdef}
\end{equation}
Direct evaluation of these Grassmann integrations lead to products of the differences between pairs of Grassmann eigenvalue functions, representing the Grassmann version of Dirac delta functionals. The definition of the Grassmann Dirac delta functional can be succinctly expressed as
\begin{equation}
\int \exp(-i q\fdia p)\ \mathcal{D}[q] = \delta[p] , ~~~~~
\int \exp(i q\fdia p)\ \mathcal{D}[p] = \delta[q] ,
\label{gdirac}
\end{equation}
in terms of Grassmann integrals over the Grassmann field variables $q$ or $p$.

With the resolutions of the identity operator, the quadrature operators can also be represented as
\begin{equation}
\eqalign{
\hat{q}_s(\mathbf{k})\mathds{I} & = \int \ket{q} q_s(\mathbf{k}) \bra{q}\ \mathcal{D}[q] , \\
\hat{p}_s(\mathbf{k})\mathds{I} & = \int \ket{p} p_s(\mathbf{k}) \bra{p}\ \mathcal{D}[p] , }
\label{defqpint}
\end{equation}
in terms of their eigenstates. These representations are useful in calculations.

\subsection{Fermion quadrature bases}

Based on the preceding discussions, we conclude that the eigenstates $\ket{q}$ and $\ket{p}$ form two mutually unbiased complete orthogonal fermion quadrature bases. Their dual spaces are produced by the fermionic adjoint
\begin{equation}
(\ket{q})^{\ddag} = \bra{q} ~~~ (\ket{p})^{\ddag} = \bra{p} .
\end{equation}
If they are parameterized by the same Grassmann parameter function and its complex conjugate, the mutually unbiased bases are related by Hermitian adjoints
\begin{equation}
(\ket{q})^{\dag} = \bra{p} ~~~ (\ket{p})^{\dag} = \bra{q} .
\end{equation}
This relationship does not have a bosonic equivalent.

\section{Grassmann Fourier analysis}

We can also interpret (\ref{diracdef}) as orthogonality conditions for exponential functionals. It allows us to define Grassmann functional Fourier integrals. The Grassmann functional Fourier transform of an arbitrary Grassmann functional $W[q]$ and its associated inverse are thus represented by
\begin{equation}
\eqalign{
\tilde{W}[p] & = \int W[q] \exp(-i q\fdia p)\ \mathcal{D}[q] , \\
W[q] & = \int \tilde{W}[p]\exp( i q\fdia p)\ \mathcal{D}[p] . }
\label{gfft}
\end{equation}
When the inverse Fourier transform is applied to the result of the Fourier transform, it reproduces the original functional.

By operating with the identity operators resolved in the opposite bases, the elements of the different bases can be represented in terms of each other
\begin{equation}
\eqalign{
\ket{q} = \mathds{I}\ket{q} & = \int \ket{p}\exp\left(-i p\fdia q\right)\ \mathcal{D}[p] , \\
\ket{p} = \mathds{I}\ket{p} & = \int \ket{q}\exp\left(i q\fdia p\right)\ \mathcal{D}[q] . }
\end{equation}
It shows that there exists a Fourier relationship between the two mutually unbiased quadrature bases. This Fourier relationship allows us to interpret their Grassmann functions $q_s(\mathbf{k})$ and $p_s(\mathbf{k})$ as the mutually orthogonal coordinates of a functional phase space.

\section{Fermionic Wigner functional}

\subsection{Definition of the Wigner functional}

In analogy with the bosonic definition, together with the knowledge of the fermion quadrature bases, we can now define the fermionic Wigner functional as
\begin{equation}
W[q,p] = \int \bra{q+\case{1}{2}x} \hat{\rho} \ket{q-\case{1}{2}x} \exp(-i x\fdia p)\ \mathcal{D}[x] .
\label{wigdef}
\end{equation}
Here $\hat{\rho}$ (nominally a fermion density operator) can represent any operator defined on the Hilbert space of all fermion states, incorporating all the spin, spatiotemporal and particle-number degrees of freedom. In \ref{wigher}, we show that the fermionic Wigner functional can equally well be defined in terms of the $p$-basis. We also show there that the Wigner functional is Hermitian self adjoint.

While the phase space of a boson field has a {\em symplectic geometry}, the phase space of fermion fields has an {\em orthogonal geometry}, due to the anticommutation relations. However, the fermion adjoint changes the situation. In \ref{symplec}, we show that the phase space represented by the fermion quadrature field variables $q$ and $p$, as defined here, has a symplectic geometry.

\subsection{Characteristic functional}

The characteristic functional is the symplectic Grassmann functional Fourier transform of the Wigner functional with respect to both $q$ and $p$. It reads
\begin{equation}
\chi[\xi,\zeta] = \int W[q,p]\exp( i\zeta\fdia p- i q\fdia\xi)\ \mathcal{D}[q,p] .
\end{equation}
The Wigner functional is recovered from the characteristic functional via the inverse symplectic Grassmann functional Fourier transform.

\subsection{Weyl transformation}

The Weyl transformation, which reproduces a fermion operator from its Wigner functional, is given by
\begin{equation}
\hat{\rho} = \int \ket{q+\case{1}{2}x} W[q,p]
 \exp( i x\fdia p) \bra{q-\case{1}{2}x}\ \mathcal{D}[p,q,x] .
\label{weyl2}
\end{equation}
In \ref{omkeerbaar}, it is shown that the Weyl transformation is the inverse of the process in (\ref{wigdef}), which calculates the Wigner functional for a fermion operator. By implication, the mapping between operators on the fermion Hilbert space and the Wigner functionals on the quadrature phase space is invertible.

It follows that the trace of an operator is represented by
\begin{equation}
\tr\{\hat{A}\} = \int W_{\hat{A}}[q,p]\ \mathcal{D}[q,p] .
\label{wigtrace}
\end{equation}
The normalization of a density operator thus implies that
\begin{equation}
\tr\{\hat{\rho}\} = \int W_{\hat{\rho}}[q,p]\ \mathcal{D}[q,p] = 1 .
\label{wignorcon}
\end{equation}

\subsection{Quadrature and ladder operators}
\label{kwadabs}

We compute the fermionic Wigner functionals for the quadrature operators as examples. Using the eigenvalue equation for the quadrature operator $\hat{q}_s(\mathbf{k})$, we get
\begin{equation}
W_{\hat{q}}[q,p] = \int \bra{q+\case{1}{2}x} \hat{q}_s(\mathbf{k})\ket{q-\case{1}{2}x} \exp(-i x\fdia p)\ \mathcal{D}[x] = q_s(\mathbf{k}) .
\end{equation}
The Wigner functional for $\hat{p}_s(\mathbf{k})$ is obtained with the aid of (\ref{binneqp}), (\ref{defqpint}) and (\ref{gdirac}):
\begin{equation}
W_{\hat{p}}[q,p] = \int \bra{q+\case{1}{2}x} \hat{p}_s(\mathbf{k})\ket{q-\case{1}{2}x} \exp(-i x\fdia p)\ \mathcal{D}[x] = p_s(\mathbf{k}) .
\end{equation}
We see that the sign of the exponent in the definition of the Wigner functional needs to be negative, otherwise we'll get $-p_s(\mathbf{k})$ instead of $p_s(\mathbf{k})$.

The Wigner functionals of linear combinations of operators are the linear combinations of the Wigner functionals of those operators. Hence, in terms of (\ref{ainqp}), the Wigner functionals of the ladder operators are
\begin{equation}
\eqalign{
W_{\hat{a}}[q,p] = \frac{1}{\sqrt{2}} \left[q_s(\mathbf{k})+ i p_s(\mathbf{k})\right] = \alpha_s(\mathbf{k}) , \\
W_{\hat{a}^{\dag}}[q,p] = \frac{1}{\sqrt{2}} \left[q_r(\mathbf{k})- i p_r(\mathbf{k})\right] \varepsilon_{r,s} = \alpha_s^*(\mathbf{k}) . }
\label{wigaac}
\end{equation}
We used (\ref{qpnaa}) to show that the Wigner functionals of the fermion ladder operators are directly given by the complex valued Grassmann parameter functions.

\section{Application: two-level fermion system}

Here, we apply the above formalism to study the evolution of a physical system where transitions between two species of fermions are mediated by a complex scalar field. The latter is represented as a classical complex-valued field in a semi-classical treatment. We first derive an evolution equation from the interaction Hamiltonian. Then we provide a solution for this evolution equation. Since the initial state in such a scenario could be an arbitrary state, the solution is represented in terms of the transformation of the field variables of such an initial state.

In the interaction picture, the Hamiltonian for the system is given by
\begin{equation}
H = \gamma\hbar\hat{a}_1^{\dag}\diamond\mathcal{T}\diamond\hat{a}_0
+\gamma\hbar\hat{a}_0^{\dag}\diamond\mathcal{T}^{\dag}\diamond\hat{a}_1 ,
\label{hamilaa}
\end{equation}
in terms of compact notation, where $\gamma$ is a coupling constant, $\hat{a}_{m,s}^{\dag}$ and $\hat{a}_{m,s}$ represent the creation and annihilation operators for the two types of fermions, denoted by subscripts $m=0,1$, respectively, and $\mathcal{T}_{r,s}$ and $\mathcal{T}_{r,s}^{\dag}$ are kernels that are derived from the classical scalar field and its complex conjugate. The ladder operators of each type of fermion satisfy the anticommutation relations in (\ref{fermiskepabs}). Those of the different types anticommute with each other. The spinor indices of the fermion operators are separately contracted with the kernels for the sake of convenience. It merely means that the scalar field is multiplied by an identity matrix in terms of spin indices. This Hamiltonian is both Hermitian and fermionic selfadjoint.

By expressing the ladder operators in terms of the fermion quadrature operators
\begin{equation}
\eqalign{
\hat{b}_{m,s}(\mathbf{k}) & = \frac{1}{\sqrt{2}} \left[\hat{q}_{m,s}(\mathbf{k})+i\hat{p}_{m,s}(\mathbf{k})\right] , \\
\hat{b}_{m,s}^{\dag}(\mathbf{k})
& = \frac{1}{\sqrt{2}} \left[\hat{q}_{m,r}(\mathbf{k})-i\hat{p}_{m,r}(\mathbf{k})\right] \varepsilon_{r,s} , }
\label{bquad}
\end{equation}
we can express the Hamiltonian in terms of fermion quadrature operators. Hence,
\begin{equation}
H = \case{1}{2} \gamma\hbar (\hat{q}_1-i \hat{p}_1)\fdia\mathcal{T}\diamond(\hat{q}_0+i \hat{p}_0)
+\case{1}{2} \gamma\hbar (\hat{q}_0-i \hat{p}_0)\fdia\mathcal{T}^{\dag}\diamond(\hat{q}_1+i \hat{p}_1) .
\label{hamil}
\end{equation}

The evolution of the fermion state is given by $i \hbar\partial_t\hat{\rho}=[H,\hat{\rho}]$, where $\hat{\rho}$ is the density operator of the fermion state. Converted into Wigner functionals, it reads
\begin{equation}
i \hbar\partial_t W_{\hat{\rho}} = W_H\star W_{\hat{\rho}} - W_{\hat{\rho}}\star W_H ,
\end{equation}
where $\star$ represents the Moyal star product, which can be represented by a Grassmann functional integral, as derived in \ref{sterprod}.

Since the ladder operators of the different types of fermions anticommute with each other, the calculation of the Wigner functional of the Hamiltonian in (\ref{hamil}) follows directly from those in section~\ref{kwadabs}. The result reads\begin{equation}
\fl W_H = \case{1}{2} \gamma\hbar (q_1-i p_1)\fdia\mathcal{T}\diamond(q_0+i p_0)
+\case{1}{2} \gamma\hbar (q_0-i p_0)\fdia\mathcal{T}^{\dag}\diamond(q_1+i p_1) ,
\label{hamilqp}
\end{equation}
where $q_{m,s}$ and $p_{m,s}$ are the Grassmann quadrature field variables.

\subsection{Evolution equation}

For the purpose of the star product calculations, we represent the Hamiltonian in (\ref{hamilqp}) in terms of a generating functional
\begin{equation}
\mathcal{G} = \exp\left(\mu_0\diamond q_0+\mu_1\diamond q_1+\nu_0\diamond p_0+\nu_1\diamond p_1\right) ,
\end{equation}
where $\mu_0$, $\mu_1$, $\nu_0$, and $\nu_1$ are auxiliary field variables, and a construction process
\begin{eqnarray}
\mathcal{C}_H & = \frac{\gamma}{2} \left(\frac{\delta}{\delta\mu_1}-i \frac{\delta}{\delta\nu_1}\right)\fdia\mathcal{T}
\diamond\left(\frac{\delta}{\delta\mu_0}+i\frac{\delta}{\delta\nu_0}\right) \nonumber \\
& +\frac{\gamma}{2}\left(\frac{\delta}{\delta\mu_0}-i \frac{\delta}{\delta\nu_0}\right)\fdia\mathcal{T}^{\dag}
\diamond\left(\frac{\delta}{\delta\mu_1}+i\frac{\delta}{\delta\nu_1}\right) .
\end{eqnarray}
The Wigner functional of the Hamiltonian is reproduced by applying the construction process to the generating functional and then setting the auxiliary field variables to zero:
\begin{equation}
W_H = \hbar\mathcal{C}_H\{\mathcal{G}\}_{\mu_0=\mu_1=\nu_0=\nu_1=0} .
\end{equation}
The evolution equation then becomes
\begin{equation}
i \partial_t W_{\hat{\rho}} = \mathcal{C}_H \left\{\mathcal{G}\star W_{\hat{\rho}}
-W_{\hat{\rho}}\star\mathcal{G}\right\}_{\mu_0=\mu_1=\nu_0=\nu_1=0} ,
\end{equation}
where we canceled the $\hbar$'s on both sides of the equation.

The next step is to compute the star products. The Grassmann functional integral, which is given in \ref{sterprod}, needs to be applied twice to address both pairs of quadrature variables. Even though the Wigner functional of the state is not specified, we can evaluate these functional integrals, because they produce Dirac delta functionals that perform replacements on the arguments of the state's Wigner functional. The results of the two star products are
\begin{equation}
\eqalign{
\mathcal{G}\star W_{\hat{\rho}}
& = \exp\left(\nu_0\diamond p_0-q_0\diamond\mu_0+\nu_1\diamond p_1-q_1\diamond \mu_1\right) \\
& \times W_{\hat{\rho}}[q_0-i\case{1}{2}\nu_0\cdot\varepsilon,p_0-i\case{1}{2}\varepsilon\cdot\mu_0,
q_1-i\case{1}{2}\nu_1\cdot\varepsilon,p_1-i\case{1}{2}\varepsilon\cdot\mu_1] , \\
W_{\hat{\rho}}\star \mathcal{G}
& = \exp\left(\nu_0\diamond p_0-q_0\diamond\mu_0+\nu_1\diamond p_1-q_1\diamond \mu_1\right) \nonumber \\
& \times W_{\hat{\rho}}[q_0+i\case{1}{2}\nu_0\cdot\varepsilon,p_0+i\case{1}{2}\varepsilon\cdot\mu_0,
q_1+i\case{1}{2}\nu_1\cdot\varepsilon,p_1+i\case{1}{2}\varepsilon\cdot\mu_1] . }
\end{equation}
We get back the product of the original generating functional and the Wigner functional of the state with displacements in terms of the auxiliary variables. The second term differs from the first only in the signs of the shifts in the arguments.

Now, we can apply the construction operation. Thus, we obtain the equation
\begin{eqnarray}
i \partial_t W_{\hat{\rho}}
& = -\frac{\gamma}{2}\left(\frac{\delta W_{\hat{\rho}}}{\delta q_1}-i \frac{\delta W_{\hat{\rho}}}{\delta p_1}\right)
\diamond\mathcal{T}\diamond (q_0+i p_0) \nonumber \\
& -\frac{\gamma}{2}(q_1-i p_1) \fdia\mathcal{T}\fdia
\left(\frac{\delta W_{\hat{\rho}}}{\delta q_0}+i \frac{\delta W_{\hat{\rho}}}{\delta p_0}\right) \nonumber \\
& -\frac{\gamma}{2}\left(\frac{\delta W_{\hat{\rho}}}{\delta q_0}-i \frac{\delta W_{\hat{\rho}}}{\delta p_0}\right)
\diamond\mathcal{T}^{\dag}\diamond (q_1+i p_1) \nonumber \\
& -\frac{\gamma}{2}(q_0-i p_0) \fdia\mathcal{T}^{\dag}\fdia
\left(\frac{\delta W_{\hat{\rho}}}{\delta q_1}+i \frac{\delta W_{\hat{\rho}}}{\delta p_1}\right) .
\end{eqnarray}
Next, we convert the quadrature field variables to complex-valued Grassmann field variables, using different Greek symbols to distinguish the two species of fermions. They are defined as
\begin{equation}
\eqalign{
\beta & = \case{1}{\sqrt{2}} (q_0+i p_0) , ~~~~~
\beta^* = \case{1}{\sqrt{2}} (q_0-i p_0)\cdot\varepsilon , \\
\zeta & = \case{1}{\sqrt{2}} (q_1+i p_1) , ~~~~~
\zeta^* = \case{1}{\sqrt{2}} (q_1-i p_1)\cdot\varepsilon . }
\label{qpnakomp}
\end{equation}
The functional derivatives are related by
\begin{equation}
\eqalign{
\frac{1}{\sqrt{2}}\left(\frac{\delta W_{\hat{\rho}}}{\delta q_0}+i \frac{\delta W_{\hat{\rho}}}{\delta p_0}\right)
& = \frac{\delta W_{\hat{\rho}}}{\delta \beta^*}\cdot\varepsilon , ~~~~~
\frac{1}{\sqrt{2}}\left(\frac{\delta W_{\hat{\rho}}}{\delta q_0}-i \frac{\delta W_{\hat{\rho}}}{\delta p_0}\right)
= \frac{\delta W_{\hat{\rho}}}{\delta \beta} , \\
\frac{1}{\sqrt{2}}\left(\frac{\delta W_{\hat{\rho}}}{\delta q_1}+i \frac{\delta W_{\hat{\rho}}}{\delta p_1}\right)
& = \frac{\delta W_{\hat{\rho}}}{\delta \zeta^*}\cdot\varepsilon , ~~~~~
\frac{1}{\sqrt{2}}\left(\frac{\delta W_{\hat{\rho}}}{\delta q_1}-i \frac{\delta W_{\hat{\rho}}}{\delta p_1}\right)
= \frac{\delta W_{\hat{\rho}}}{\delta \zeta} . }
\end{equation}
In terms of complex-valued Grassmann field variables, the evolution equation then reads
\begin{equation}
\fl i \partial_t W_{\hat{\rho}}
= \gamma\beta^*\diamond\mathcal{T}^{\dag}\diamond\frac{\delta W_{\hat{\rho}}}{\delta\zeta^*}
-\gamma\frac{\delta W_{\hat{\rho}}}{\delta\zeta} \diamond\mathcal{T}\diamond\beta
+\gamma\zeta^*\diamond\mathcal{T}\diamond\frac{\delta W_{\hat{\rho}}}{\delta\beta^*}
-\gamma\frac{\delta W_{\hat{\rho}}}{\delta\beta} \diamond\mathcal{T}^{\dag}\diamond\zeta ,
\label{evolvgl}
\end{equation}
where two of the terms became positive because the spin transformation matrix flipped to the other side. The resulting equation is anti-Hermitian on both sides of the equation. It is a functional Fokker-Planck equation, but without dissipative terms (second order functional derivatives). As a result, it represents pure unitary evolution, maintaining the purity of the initial state.

\subsection{Solution}

Due to its unitary nature, the functional Fokker-Planck equation in (\ref{evolvgl}) admits solutions represented by transformations of the Wigner functional arguments of an arbitrary initial state. Since the equation replaces one type of field variable by another field variable, the transformation is expect to be represented in the form of linear combinations of both field variables, given by
\begin{equation}
\eqalign{
\beta \rightarrow & A(t)\diamond\beta+B(t)\diamond\zeta = \bar{\beta} , ~~~
\beta^* \rightarrow \beta^*\diamond A^{\dag}(t)+\zeta^*\diamond B^{\dag}(t) = \bar{\beta}^* , \\
\zeta \rightarrow & C(t)\diamond\zeta+D(t)\diamond\beta = \bar{\zeta} , ~~~
\zeta^* \rightarrow \zeta^*\diamond C^{\dag}(t)+\beta^*\diamond D^{\dag}(t) = \bar{\zeta}^* , }
\end{equation}
where $A(t)$, $B(t)$, $C(t)$, and $D(t)$ are unknown kernels to be solved. The transformed initial state, which solves the equation in (\ref{evolvgl}), is then given by
\begin{eqnarray}
W_{{\rm in}} \rightarrow W_{\hat{\rho}}
= & W_{{\rm in}}[\beta^*\diamond A^{\dag}+\zeta^*\diamond B^{\dag}, A\diamond\beta+B\diamond\zeta, \nonumber \\
& \zeta^*\diamond C^{\dag}+\beta^*\diamond D^{\dag}, C\diamond\zeta+D\diamond\beta] .
\label{trainit}
\end{eqnarray}
It remains to solve the four unknown kernels. The time derivative of the transformed initial state produces
\begin{eqnarray}
\partial_t W_{\hat{\rho}} & = \left[\beta^*\diamond \partial_t A^{\dag}+\zeta^*\diamond \partial_t B^{\dag}\right]
\diamond\frac{\delta W_{{\rm in}}}{\delta \bar{\beta}^*} \nonumber \\
& + \frac{\delta W_{{\rm in}}}{\delta \bar{\beta}}
\diamond\left[(\partial_t A)\diamond\beta+(\partial_t B)\diamond\zeta\right] \nonumber \\
& + \left[\zeta^*\diamond \partial_t C^{\dag}+\beta^*\diamond \partial_t D^{\dag}\right]
\diamond \frac{\delta W_{{\rm in}}}{\delta \bar{\zeta}^*} \nonumber \\
& + \frac{\delta W_{{\rm in}}}{\delta \bar{\zeta}}
\diamond\left[(\partial_t C)\diamond\zeta+(\partial_t D)\diamond\beta\right] .
\end{eqnarray}
The functional derivatives of the transformed initial state are
\begin{equation}
\eqalign{
\frac{\delta W_{\hat{\rho}}}{\delta\beta^*}
= A^{\dag}\diamond\frac{\delta W_{{\rm in}}}{\delta \bar{\beta}^*}
+D^{\dag}\diamond\frac{\delta W_{{\rm in}}}{\delta \bar{\zeta}^*} , ~~~
\frac{\delta W_{\hat{\rho}}}{\delta\beta}
= \frac{\delta W_{{\rm in}}}{\delta \bar{\beta}}\diamond A
+\frac{\delta W_{{\rm in}}}{\delta\bar{\zeta}}\diamond D , \\
\frac{\delta W_{\hat{\rho}}}{\delta\zeta^*}
= C^{\dag}\diamond\frac{\delta W_{{\rm in}}}{\delta \bar{\zeta}^*}
+B^{\dag}\diamond\frac{\delta W_{{\rm in}}}{\delta \bar{\beta}^*} , ~~~
\frac{\delta W_{\hat{\rho}}}{\delta\zeta}
= \frac{\delta W_{{\rm in}}}{\delta \bar{\zeta}}\diamond C
+\frac{\delta W_{{\rm in}}}{\delta\bar{\beta}}\diamond B . }
\end{equation}
Substituting the time derivative and functional derivatives into (\ref{evolvgl}), we obtain an equation that relates the time derivatives of the kernels with contractions of these kernels by $\mathcal{T}$ and $\mathcal{T}^{\dag}$. Comparing the terms on both sides of the equation in terms of the field variables and functional derivatives, we can extract eight equations for the kernels, being
\begin{equation}
\eqalign{
\partial_t A & = i\gamma B\diamond\mathcal{T} , ~~~~~
\partial_t B = i\gamma A\diamond\mathcal{T}^{\dag} , \\
\partial_t C & = i\gamma D\diamond\mathcal{T}^{\dag} , ~~~~~
\partial_t D = i\gamma C\diamond\mathcal{T} , }
\end{equation}
together with their Hermitian conjugates. As a result, we have two pairs of equations, with which we can solve $A$ and $B$, and $C$ and $D$, respectively. These sets are readily solved in the normal way by integration and repeated back-substitution, with initial conditions $A(0)=C(0)=\mathbf{1}$ and $B(0)=D(0)=0$, leading to
\begin{equation}
\eqalign{
\fl A(t) = \mathbf{1}-\gamma^2 \int_0^t \int_0^{t_1} \mathcal{T}^{\dag}(t_2)\diamond\mathcal{T}(t_1)\ dt_2\ dt_1 \\
\fl \ \ \ +\gamma^4 \int_0^t \int_0^{t_1} \int_0^{t_2} \int_0^{t_3} \mathcal{T}^{\dag}(t_4)\diamond\mathcal{T}(t_3)
\diamond\mathcal{T}^{\dag}(t_2)\diamond\mathcal{T}(t_1)\ dt_4\ dt_3\ dt_2\ dt_1 + ...\ , \\
\fl B(t) = i\gamma \int_0^t \mathcal{T}^{\dag}(t_1)\ dt_1
- i \gamma^3 \int_0^t \int_0^{t_1} \int_0^{t_2} \mathcal{T}^{\dag}(t_3)\diamond\mathcal{T}(t_2)
\diamond\mathcal{T}^{\dag}(t_1)\ dt_3\ dt_2\ dt_1 + ...\ , \\
\fl C(t) = \mathbf{1}-\gamma^2 \int_0^t \int_0^{t_1} \mathcal{T}(t_2)\diamond\mathcal{T}^{\dag}(t_1)\ dt_2\ dt_1 \\
\fl \ \ \ +\gamma^4 \int_0^t \int_0^{t_1} \int_0^{t_2} \int_0^{t_3} \mathcal{T}(t_4)\diamond\mathcal{T}^{\dag}(t_3)
\diamond\mathcal{T}(t_2)\diamond\mathcal{T}^{\dag}(t_1)\ dt_4\ dt_3\ dt_2\ dt_1 + ...\ , \\
\fl D(t) = i\gamma \int_0^t \mathcal{T}(t_1)\ dt_1
- i \gamma^3 \int_0^t \int_0^{t_1} \int_0^{t_2} \mathcal{T}(t_3)\diamond\mathcal{T}^{\dag}(t_2)
\diamond\mathcal{T}(t_1)\ dt_3\ dt_2\ dt_1 + ...\ . }
\label{oplkerne}
\end{equation}
These expressions, together with (\ref{trainit}) represent the solutions for the functional Fokker-Planck equation in (\ref{evolvgl}) for an arbitrary initial state.

Considering the expressions in (\ref{oplkerne}), we see that $A$ and $C$ consist of all the even contractions of $\mathcal{T}$ and $\mathcal{T}^{\dag}$, while $B$ and $D$ contain all the odd contractions of $\mathcal{T}$ and $\mathcal{T}^{\dag}$. It thus follows that, in scenarios where $\mathcal{T}$ and $\mathcal{T}^{\dag}$ are represented by constants, the kernels in (\ref{oplkerne}) become simple trigonometric functions, representing Rabi oscillations of the two species of fermions. The more general case, as represented by (\ref{oplkerne}), thus generalizes the notion of Rabi oscillations for cases with more complicated spatiotemporal properties.

\section{Conclusions}

A fermionic Wigner functional formalism is developed for fermions on a Grassmann phase space with the aid of fermion quadrature bases. Much of the development involves the identification of suitable quadrature bases as eigenstates of suitable fermion quadrature operators. The Majorana operators are not suitable, because their properties differ too much from their bosonic counterparts. Instead, we find that certain fermion Bogoliubov operators are analogues to boson quadrature operators, provided that the Hermitian adjoint incorporates a spin transformation, which we call the fermionic adjoint. As a result, these fermion quadrature operators are not Hermitian, unlike their bosonic counterparts, and thus do not represent observables.

The eigenstates of the fermion Bogoliubov operators, which are displaced squeezed vacuum states, are obtained with the aid of the algebra of Grassmann even operators, constructed from the fermion ladder operators and complex Grassmann parameter functions. Each eigenstate is associated with a Grassmann eigenvalue function that carries spin and spatiotemporal degrees of freedom.

We select suitable fermion quadrature operators as those fermion Bogoliubov operators that are fermionic selfadjoint. Their left-eigenstates are the fermionic duals of their right-eigenstates. The eigenstates are similar to their bosonic counterparts, but with some differences. The most important difference is the fact that the dual space is not given by the Hermitian adjoint, but by the fermionic adjoint. The eigenstates do not satisfy orthogonality relationships with their Hermitian adjoints. However, the left- and right-eigenstates of each operator, which are related via the fermionic adjoint, do satisfy orthogonality relationships. As such, these eigenstates serve as complete orthogonal fermion quadrature bases. Their associated eigenvalue functions are self-conjugate; they remain unchanged under fermionic conjugation. The eigenvalue functions eventually serve as field variables of the Grassmann functional phase space. The latter has a symplectic structure, which is usually associated with bosonic phase spaces.

The completeness relations for the fermion quadrature bases, together with their orthogonality conditions, lead to a Fourier relationship between the two mutually unbiased quadrature bases. It allows for the definition of a functional Fourier relationship that transforms functionals of one quadrature field variable into those of the other quadrature field variable.

The fermionic Wigner functional is now defined with the aid of the quadrature basis and the functional Fourier transform. It can represent any operator on the Hilbert space of the fermions, incorporating all the spatiotemporal and spin degrees of freedom with the particle-number degrees of freedom. The characteristic functional is given by the symplectic functional Fourier transform of a fermionic Wigner functional with respect to both field variables. The associated Weyl transformation converts fermionic Wigner functionals back into operators.

In the end, we apply the fermionic Wigner functional theory to an application in which the transition between two species of fermions is mediated by a classical complex scalar field. We represent the dynamics by a semi-classical interaction term. After deriving the functional Fokker-Planck equation for the process, we obtain solutions in the form of transformed initial states. These solutions provide a generalization of the Rabi oscillations found in simple scenarios of this kind.

We do not address the issue of the calculation of measurements on Grassmann phase spaces here. Such a treatment needs a careful presentation, which we plan to address in some future publication. A brief discussion of some aspects of measurements on Grassmann phase spaces can be found in \cite{caglaub}.


\appendix

\section{Algebra of Grassmann even operators}
\label{boscomm}

The complete algebra of Grassmann even operators obeys the following commutation relations. Those among spectral operators are
\begin{equation}
\eqalign{
\left[\hat{\alpha}, \hat{\beta}^{\dag}\right] = \alpha^*\diamond\beta , ~~~
\left[\hat{\alpha}_{\varepsilon}, \hat{\beta}_{\varepsilon}^{\dag}\right] = -\beta^*\diamond\alpha , \\
\left[\hat{\alpha}_{\varepsilon}, \hat{\beta}^{\dag}\right] = \alpha\fdia\beta , ~~~
\left[\hat{\alpha}, \hat{\beta}_{\varepsilon}^{\dag}\right] = \alpha^*\fdia\beta^* , }
\end{equation}
where we define the spectral operators as
\begin{equation}
\eqalign{
\hat{\alpha} = \alpha^*\diamond\hat{a} , ~~~
\hat{\beta}^{\dag} = \hat{a}^{\dag}\diamond\beta , \\
\hat{\alpha}_{\varepsilon} = \alpha\fdia\hat{a} , ~~~
\hat{\beta}_{\varepsilon}^{\dag} = \hat{a}^{\dag}\fdia\beta^* , }
\end{equation}
with Grassmann spectral functions $\alpha$, $\beta$, $\alpha^*$, and $\beta^*$. Those involving $\hat{R}$ and $\hat{R}^{\dag}$ are
\begin{equation}
\eqalign{
\left[\hat{\alpha}, \hat{R}^{\dag}\right] = \hat{\alpha}_{\varepsilon}^{\dag} , ~~~
\left[\hat{\alpha}_{\varepsilon}, \hat{R}^{\dag}\right] = - \hat{\alpha}^{\dag} , \\
\left[\hat{R}, \hat{\alpha}^{\dag}\right] = \hat{\alpha}_{\varepsilon} , ~~~
\left[\hat{R}, \hat{\alpha}_{\varepsilon}^{\dag}\right] = - \hat{\alpha} , ~~~
\left[\hat{R}, \hat{R}^{\dag}\right] = -\hat{s} , }
\end{equation}
where $\hat{s}$ is a symmetrized number operator
\begin{equation}
\hat{s} = \case{1}{2} \int \hat{a}_s^{\dag}(\mathbf{k}) \hat{a}_s(\mathbf{k})
- \hat{a}_s(\mathbf{k}) \hat{a}_s^{\dag}(\mathbf{k})\ {\rm d}k_E
= \hat{a}^{\dag}\diamond\hat{a} - \case{1}{2} \Omega ,
\label{numr}
\end{equation}
with the divergent constant $\Omega$, given by
\begin{equation}
\Omega = \int \delta(0)\ {\rm d}^3 k .
\label{defomega}
\end{equation}
The commutation relations that involve $\hat{s}$ are
\begin{equation}
\eqalign{
\left[\hat{\alpha}, \hat{s}\right] = \hat{\alpha} , ~~~
\left[\hat{\alpha}_{\varepsilon}, \hat{s}\right] = \hat{\alpha}_{\varepsilon} , ~~~
\left[\hat{R}, \hat{s}\right] = 2\hat{R} , \\
\left[\hat{s}, \hat{\alpha}^{\dag}\right] = \hat{\alpha}^{\dag} , ~~~
\left[\hat{s}, \hat{\alpha}_{\varepsilon}^{\dag}\right] = \hat{\alpha}_{\varepsilon}^{\dag} , ~~~
\left[\hat{s}, \hat{R}^{\dag}\right] = 2\hat{R}^{\dag} . }
\end{equation}
The subset of operators without Grassmann spectral function $\hat{R}$, $\hat{R}^{\dag}$, and $\hat{s}$, form an algebra for SU(1,1).

\section{Products with exponentiated operators}
\label{boseksop}

Here, we use the identity in (\ref{eksopprod}) to compute the products of exponentiated operators. Those involving only spectral operators are
\begin{equation}
\eqalign{
\exp(c\hat{\alpha})\hat{\beta}^{\dag}\exp(-c\hat{\alpha}) = \hat{\beta}^{\dag} + c \alpha^*\diamond \beta , \\
\exp(c\hat{\alpha}_{\varepsilon})\hat{\beta}^{\dag}\exp(-c\hat{\alpha}_{\varepsilon}) = \hat{\beta}^{\dag} + c \alpha\fdia \beta , \\
\exp(c\hat{\alpha})\hat{\beta}_{\varepsilon}^{\dag}\exp(-c\hat{\alpha}) = \hat{\beta}_{\varepsilon}^{\dag} + c \alpha^*\fdia \beta^* , \\
\exp(c\hat{\alpha}_{\varepsilon})\hat{\beta}_{\varepsilon}^{\dag}\exp(-c\hat{\alpha}_{\varepsilon})
= \hat{\beta}_{\varepsilon}^{\dag} - c \beta^*\diamond \alpha , }
\end{equation}
where $c$ is an arbitrary complex c-number. Including $\hat{R}$ and $\hat{R}^{\dag}$, we have
\begin{equation}
\eqalign{
\exp(c\hat{R})\hat{\alpha}^{\dag}\exp(-c\hat{R}) = \hat{\alpha}^{\dag} + c\hat{\alpha}_{\varepsilon} , \\
\exp(c\hat{\alpha})\hat{R}^{\dag}\exp(-c\hat{\alpha})
= \hat{R}^{\dag} + c\hat{\alpha}_{\varepsilon}^{\dag} + \case{1}{2} c^2 \alpha^*\fdia \alpha^* , \\
\exp(c\hat{R})\hat{\alpha}_{\varepsilon}^{\dag}\exp(-c\hat{R}) = \hat{\alpha}_{\varepsilon}^{\dag} - c\hat{\alpha} , \\
\exp(c\hat{\alpha}_{\varepsilon})\hat{R}^{\dag}\exp(-c\hat{\alpha}_{\varepsilon})
= \hat{R}^{\dag} - c\hat{\alpha}^{\dag} - \case{1}{2} c^2 \alpha\fdia \alpha , \\
\exp(c\hat{R})\hat{R}^{\dag}\exp(-c\hat{R}) = \hat{R}^{\dag} - c\hat{s} - c^2\hat{R} , }
\end{equation}
and then with $\hat{s}$, we have
\begin{equation}
\eqalign{
\exp(c\hat{\alpha})\hat{s}\exp(-c\hat{\alpha}) = \hat{s} +c\hat{\alpha} , ~~~
\exp(c\hat{\alpha}_{\varepsilon})\hat{s}\exp(-c\hat{\alpha}_{\varepsilon}) = \hat{s} +c\hat{\alpha}_{\varepsilon} , \\
\exp(c\hat{s})\hat{\alpha}^{\dag}\exp(-c\hat{s}) = \exp(c) \hat{\alpha}^{\dag} , ~~~
\exp(c\hat{s})\hat{\alpha}_{\varepsilon}^{\dag}\exp(-c\hat{s}) = \exp(c) \hat{\alpha}_{\varepsilon}^{\dag} , \\
\exp(c\hat{R})\hat{s}\exp(-c\hat{R}) = \hat{s} +2c\hat{R} , ~~~
\exp(c\hat{s})\hat{R}^{\dag}\exp(-c\hat{s}) = \exp(2c) \hat{R}^{\dag} . }
\label{eksops}
\end{equation}

In addition to these generic identities, we need a few special identities to simplify the left-hand side of the inner product equation, for which we need additional generic commutation relations. They are
\begin{equation}
\eqalign{
[\xi\hat{R}^{\dag}-\xi^*\hat{R},c^*\hat{\alpha}^{\dag}-c\hat{\alpha}]
= \xi c\hat{\alpha}_{\varepsilon}^{\dag}-\xi^* c^*\hat{\alpha}_{\varepsilon} , \\
[\xi\hat{R}^{\dag}-\xi^*\hat{R},\zeta\hat{R}^{\dag}-\zeta^*\hat{R}]
= (\xi^*\zeta-\zeta^*\xi) \hat{s} , \\
[\xi\hat{R}^{\dag}-\xi^*\hat{R},c^*\hat{\alpha}_{\varepsilon}-c\hat{\alpha}_{\varepsilon}^{\dag}]
= \xi c^*\hat{\eta}^{\dag}-\xi^*c\hat{\eta} , \\
[\xi\hat{R}^{\dag}-\xi^*\hat{R},\hat{s}]
= -2(\xi^*\hat{R}+\xi\hat{R}^{\dag}) , \\
[c_1^*\hat{\alpha}^{\dag}-c_1\hat{\alpha},c_2^*\hat{\alpha}_{\varepsilon}-c_2\hat{\alpha}_{\varepsilon}^{\dag}]
= c_1 c_2 \alpha^*\fdia\alpha^*-c_1^* c_2^* \alpha\fdia\alpha , }
\end{equation}
where $c$, $c_1$, and $c_2$ are arbitrary complex c-numbers. Note that some cases produce a cyclic behavior with one sign change. These commutation relations aid the expansions for the products of exponentiated operators found on the left-hand side of (\ref{omksdt}):
\begin{equation}
\eqalign{
\exp\left(-t\xi\hat{R}^{\dag}+t\xi^*\hat{R}\right)
\left(\hat{\eta}^{\dag}-\hat{\eta}\right)\exp\left(t\xi\hat{R}^{\dag}-t\xi^*\hat{R}\right) \\
= \cos(t|\xi|) (\hat{\eta}^{\dag}-\hat{\eta})
- \sin(t|\xi|) (\Phi\hat{\eta}_{\varepsilon}^{\dag}-\Phi^*\hat{\eta}_{\varepsilon}) , \\
\exp\left(t\hat{\eta}^{\dag}-t\hat{\eta}\right)\left(\zeta\hat{R}^{\dag}-\zeta^*\hat{R}\right)
\exp\left(-t\hat{\eta}^{\dag}+t\hat{\eta}\right) \\
= \zeta\hat{R}^{\dag}-\zeta^*\hat{R} +t (\zeta^*\hat{\eta}_{\varepsilon}-\zeta\hat{\eta}_{\varepsilon}^{\dag})
 + \case{1}{2} t^2 (\zeta\eta^*\fdia\eta^*-\zeta^*\eta\fdia\eta) , \\
\exp\left(-t\xi\hat{R}^{\dag}+t\xi^*\hat{R}\right)\left(\zeta\hat{R}^{\dag}-\zeta^*\hat{R}\right)
\exp\left(t\xi\hat{R}^{\dag}-t\xi^*\hat{R}\right) \\
= \zeta\hat{R}^{\dag}-\zeta^*\hat{R} - \case{1}{2} \sin(2t|\xi|) (\Phi^*\zeta-\zeta^*\Phi) \hat{s} \\
\ \ - \case{1}{2} [1-\cos(2t|\xi|)] (\Phi^*\zeta-\zeta^*\Phi) (\Phi^*\hat{R}+\Phi\hat{R}^{\dag}) , \\
\exp\left(-t\xi\hat{R}^{\dag}+t\xi^*\hat{R}\right)\left(\zeta^*\hat{\eta}_{\varepsilon}-\zeta\hat{\eta}_{\varepsilon}^{\dag}\right)
\exp\left(t\xi\hat{R}^{\dag}-t\xi^*\hat{R}\right) \\
= \cos(t|\xi|) \left(\zeta^*\hat{\eta}_{\varepsilon}-\zeta\hat{\eta}_{\varepsilon}^{\dag}\right)
+ \sin(t|\xi|) \left(\Phi^*\zeta\hat{\eta}-\Phi\zeta^*\hat{\eta}^{\dag}\right) . }
\end{equation}
where we defined $\xi=|\xi|\Phi$.

\section{Hermitian and fermionic adjoints}
\label{herfer}

Various Grassmann even combinations represented as contractions among ladder operators and Grassmann fields are found in the analyses. Sometimes, their Hermitian or fermionic adjoints need to be computed. Here, we provide a list of all such transformations. The Hermitian adjoint of all $\diamond$-contractions are directly obtained as
\begin{equation}
(\alpha^*\diamond\hat{a})^{\dag} = \hat{a}^{\dag}\diamond\alpha , ~~~
(\hat{a}^{\dag}\diamond\alpha)^{\dag} = \alpha^*\diamond\hat{a} , ~~~
(\alpha^*\diamond\alpha)^{\dag} = \alpha^*\diamond\alpha .
\end{equation}
With the $\fdia$-contractions we need to use the properties given in (\ref{epsprop}), to obtain
\begin{equation}
\eqalign{
(\hat{a}^{\dag}\fdia\alpha^*)^{\dag} & = \alpha\fdia\hat{a} , ~~~
(\alpha\fdia\hat{a})^{\dag} = \hat{a}^{\dag}\fdia\alpha^* , \\
(\alpha^*\fdia\alpha^*)^{\dag} & = \alpha\fdia\alpha , ~~~
(\alpha\fdia\alpha)^{\dag} = \alpha^*\fdia\alpha^* . }
\end{equation}

For the fermionic adjoints, we use (\ref{feradj}) and (\ref{ferconj}). Those with $\diamond$-contractions produce
\begin{equation}
(\alpha^*\diamond\hat{a})^{\ddag} = \hat{a}^{\dag}\diamond\alpha , ~~~
(\hat{a}^{\dag}\diamond\alpha)^{\ddag} = \alpha^*\diamond\hat{a} , ~~~
(\alpha^*\diamond\alpha)^{\ddag} = \alpha^*\diamond\alpha ,
\end{equation}
and for the $\fdia$-contractions, we have
\begin{equation}
\eqalign{
(\hat{a}^{\dag}\fdia\alpha^*)^{\ddag} & = - \alpha\fdia\hat{a} , ~~~
(\alpha\fdia\hat{a})^{\ddag} = - \hat{a}^{\dag}\fdia\alpha^* , \\
(\alpha^*\fdia\alpha^*)^{\ddag} & = - \alpha\fdia\alpha , ~~~
(\alpha\fdia\alpha)^{\ddag} = - \alpha^*\fdia\alpha^* . }
\end{equation}
The sign change comes from the $\alpha$ or $\alpha^*$ on the ``wrong'' side of the contraction, leading to an $\varepsilon$ that needs to be transposed.

When the contractions involve quadrature field variables, we need to use their definitions in terms of $\alpha$ and $\alpha^*$, as given in (\ref{eiewaar}). First, we note that
\begin{equation}
q_s^* = -i p_r \varepsilon_{r,s} , ~~~
p_s^* = i q_r \varepsilon_{r,s} .
\end{equation}
Hence, if they have the same parameter functions, we get
\begin{equation}
q^*\diamond q = -i p\fdia q = -i q\fdia p , ~~~
p^*\diamond p = i q\fdia p .
\end{equation}
The Hermitian adjoints for different contractions are
\begin{equation}
\eqalign{
(q_1\diamond q_2)^{\dag} & = q_2^*\diamond q_1^* = p_2\diamond p_1 , ~~~
(q_1\diamond p_2)^{\dag} = p_2^*\diamond q_1^* = - q_2\diamond p_1 , \\
(q_1\fdia q_2)^{\dag} & = q_2^*\fdia q_1^* = p_2\fdia p_1 , ~~~
(q_1\fdia p_2)^{\dag} = p_2^*\fdia q_1^* = - q_2\fdia p_1 . }
\label{hadjeie}
\end{equation}
Since $q^{\ddag}=q$ and $p^{\ddag}=p$, the fermionic adjoints of these contractions give
\begin{equation}
\eqalign{
(q_1\diamond q_2)^{\ddag} & = q_2\diamond q_1 = -q_1\diamond q_2 , ~~~
(q_1\fdia q_2)^{\ddag} = q_2\fdia q_1 = q_1\fdia q_2 , \\
(p_1\diamond p_2)^{\ddag} & = p_2\diamond p_1 = -p_1\diamond p_2 , ~~~
(p_1\fdia p_2)^{\ddag} = p_2\fdia p_1 = p_1\fdia p_2 . }
\label{fadjeie}
\end{equation}

\section{Hermitian adjoint of a Wigner functional}
\label{wigher}

First, we show that the fermionic Wigner functional can be defined in terms of the $p$-basis. To this end, we insert identities resolved in terms of $p$-bases, as in (\ref{volqp}), on either side of the operator in (\ref{wigdef}). A change of variables performed on the two $p$-variables of these resolved identities, leads to
\begin{eqnarray}
W[q,p] & = \int \exp\left(i q\fdia p_1+i \case{1}{2}x\fdia p_1\right) \bra{p_1}\hat{\rho}\ket{p_2}
\exp\left(-i q\fdia p_2+i \case{1}{2}x\fdia p_2\right) \nonumber \\
& \times \exp(-i x\fdia p)\ \mathcal{D}[x]\ \mathcal{D}[p_1,p_2] \nonumber \\
& = \int \bra{p_0+\case{1}{2}y}\hat{\rho}\ket{p_0-\case{1}{2}y} \exp(i q\fdia y) \nonumber \\
& \times  \exp(-i x\fdia p+i x\fdia p_0)\ \mathcal{D}[x]\ \mathcal{D}[p_0,y] \nonumber \\
& = \int \bra{p+\case{1}{2}y}\hat{\rho}\ket{p-\case{1}{2}y} \exp\left(i q\fdia y\right)\ \mathcal{D}[y] .
\label{wigdefp}
\end{eqnarray}
Note that, while $x$ is a $q$-type field variable, $y$ is a $p$-type field variable. Next we perform a Hermitian adjoint on (\ref{wigdef}). It produces
\begin{equation}
W^{\dag}[q,p] = \int \bra{p-\case{1}{2}y'} \hat{\rho}^{\dag} \ket{p+\case{1}{2}y'} \exp(-i q\fdia y')\ \mathcal{D}[y'] .
\end{equation}
where we used (\ref{hadjeie}), and
\begin{equation}
\left(\bra{q+\case{1}{2}x}\right)^{\dag} = \ket{p+\case{1}{2}y'} , ~~~
\left(\ket{q-\case{1}{2}x}\right)^{\dag} = \bra{p-\case{1}{2}y'} .
\end{equation}
As a final step, we redefine $y'\rightarrow -y$. Since it reproduces (\ref{wigdefp}), when $\hat{\rho}^{\dag}=\hat{\rho}$, it shows that $W^{\dag}[q,p]=W[q,p]$ for the Wigner functional of a state.

\section{Symplectic nature of the field variables}
\label{symplec}

One can generalize the Bogoliubov transformation in (\ref{ferbog}) by allowing the constants $u$ and $v$ to become suitable Bogoliubov kernels with their spin structures given by Kronecker deltas: $U_{r,s}(\mathbf{k},\mathbf{k}')=\delta_{r,s}U(\mathbf{k},\mathbf{k}')$ and $V_{r,s}(\mathbf{k},\mathbf{k}') =\delta_{r,s}V(\mathbf{k},\mathbf{k}')$. The Bogoliubov transformation then becomes
\begin{equation}
\eqalign{
\hat{b} & = U\diamond\hat{a}-V\fdia\hat{a}^{\dag} = U\diamond\hat{a}+\hat{a}^{\dag}\fdia V^T  , \\
\hat{b}^{\dag} & = \hat{a}^{\dag}\diamond U^{\dag}-\hat{a}\fdia V^{\dag} = \hat{a}^{\dag}\diamond U^{\dag}+V^*\fdia\hat{a} , }
\label{genferbog}
\end{equation}
and the inverse Bogoliubov transformation is
\begin{equation}
\eqalign{
\hat{a} & = U\diamond\hat{b}+V\fdia\hat{b}^{\dag} = U\diamond\hat{b}-\hat{b}^{\dag}\fdia V^T , \\
\hat{a}^{\dag} & = \hat{b}^{\dag}\diamond U^{\dag}+\hat{b}\fdia V^{\dag} = \hat{b}^{\dag}\diamond U^{\dag}-V^*\fdia\hat{b} . }
\end{equation}
To satisfy the anticommutation relations for such Bogoliubov operators, and the inverse transformation, the kernels have the properties that $U=U^{\dag}$, $V=V^T$, and
\begin{equation}
U\diamond U+V\diamond V^* = \mathbf{1} , ~~~
U\diamond V-V\diamond U^* = 0 .
\label{uvid}
\end{equation}
By applying (\ref{wigdef}) to the generalized Bogoliubov transformation in (\ref{genferbog}), using (\ref{wigaac}), we obtain an equivalent Bogoliubov transformation that applies to the complex-valued Grassmann field variables (i.e., Grassmann parameter functions), given by
\begin{equation}
\eqalign{
\beta & = U\diamond\alpha-V\fdia\alpha^* = U\diamond\alpha+\alpha^*\fdia V , \\
\beta^* & = \alpha^*\diamond U-\alpha\fdia V^* = \alpha^*\diamond U+V^*\fdia\alpha . }
\label{fvtra}
\end{equation}
The complex-valued field variables $\alpha$ and $\alpha^*$ are represented in terms of the quadrature field variables by (\ref{qpnaa}). For the transformed complex-valued field variables $\beta$ and $\beta^*$, an equivalent representation is given by
\begin{equation}
\beta = \case{1}{\sqrt{2}} (g+i h) , ~~~~~
\beta^* = \case{1}{\sqrt{2}} (g-i h)\cdot\varepsilon ,
\label{ghnab}
\end{equation}
where $g$ and $h$ are the transformed quadrature field variables, associated with $\beta$ and $\beta^*$. By substituting (\ref{qpnaa}) and (\ref{ghnab}) into (\ref{fvtra}), we obtain the canonical transformation between sets of quadrature field variables. As a matrix-vector equation, it reads
\begin{equation}
\mathbf{g} = M \mathbf{q} ,
\end{equation}
where $\mathbf{g}=[g~h]^T$, $\mathbf{q}=[q~p]^T$, and
\begin{equation}
M = \left[ \begin{array}{cc} \case{1}{2}(U+U^*+V-V^*) & i\case{1}{2}(U-U^*-V-V^*) \\
-i\case{1}{2}(U-U^*+V+V^*) & \case{1}{2}(U+U^*-V+V^*) \end{array} \right] ,
\end{equation}
is the canonical transformation matrix. Using (\ref{uvid}), we can now show that $M$ is a {\em symplectic matrix}, satisfying the equation $M J M^T = J$, \cite{goldstein9} where
\begin{equation}
J = \left[ \begin{array}{cc} 0 & \mathbf{1} \\ -\mathbf{1} & 0 \end{array} \right] .
\end{equation}
It demonstrates that the Grassmann phase space, defined in terms of the quadrature field variables $q$ and $p$ has a symplectic geometry.

\section{Invertibility of the Wigner mapping}
\label{omkeerbaar}

The Weyl transformation in (\ref{weyl2}) is presented as the inverse of Wigner functional calculation in (\ref{wigdef}). To show that this mapping between operators on the fermion Hilbert space and the Wigner functionals on the quadrature phase space is invertible, we substitute (\ref{wigdef}) into (\ref{weyl2}). It gives
\begin{eqnarray}
\hat{\rho}' & = \int \ket{q+\case{1}{2}x} \bra{q+\case{1}{2}x'} \hat{\rho} \ket{q-\case{1}{2}x'} \bra{q-\case{1}{2}x} \nonumber \\
& \times \exp( i x\fdia p-i x'\fdia p)\ \mathcal{D}[p,q,x,x']  \nonumber \\
& = \int \ket{q+\case{1}{2}x} \bra{q+\case{1}{2}x} \hat{\rho} \ket{q-\case{1}{2}x} \bra{q-\case{1}{2}x}\ \mathcal{D}[q,x] .
\end{eqnarray}
For the last expression, we first evaluated the integral over $p$ to produce a Dirac delta function $\delta[x-x']$. Then we evaluated the integral over $x'$. The expression relates the original operator $\hat{\rho}$ back to an operator $\hat{\rho}'$. The aim is to show that $\hat{\rho}'=\hat{\rho}$. Next, we perform a redefinition of the integration field variables so that $q+\case{1}{2}x\rightarrow q_1$ and $q-\case{1}{2}x\rightarrow q_2$. The result then leads to identity operators, so that
\begin{equation}
\hat{\rho}' = \int \ket{q_1} \bra{q_1} \hat{\rho} \ket{q_2} \bra{q_2}\ \mathcal{D}[q_1,q_2]
= \mathds{I} \hat{\rho} \mathds{I} = \hat{\rho} .
\end{equation}
It demonstrates that the Weyl transformation in (\ref{weyl2}) is indeed the inverse of Wigner functional calculation in (\ref{wigdef}).

\section{Grassmann star products}
\label{sterprod}

The Wigner functional for the product of two operators, each of which is expressed with the aid of (\ref{weyl2}) as a Weyl transformation of a Wigner functional, can be represented by a Grassmann functional integral. It represents that {\em Grassmann star product}, given by
\begin{eqnarray}
W_{\hat{A}\hat{B}}[q,p] & = \int \delta\left[q+\case{1}{2}x-q_a\right] W_{\hat{A}}\left[\case{1}{2}q_a+\case{1}{2}q_b,p_1\right]
\exp[ i (q_a-q_b)\fdia p_1] \nonumber \\
& \delta\left[q_b-q_c\right] W_{\hat{B}}\left[\case{1}{2}q_c+\case{1}{2}q_d,p_2\right] \exp[ i (q_c-q_d)\fdia p_2] \nonumber \\
& \times \delta\left[q_d-q+\case{1}{2}x\right] \exp(- i x\fdia p)\ \mathcal{D}[p_1,p_2,q_a,q_b,q_c,q_d,x] \nonumber \\
& = \mathcal{N}_0^{-2} \int W_{\hat{A}}[q_1,p_1] W_{\hat{B}}[q_2,p_2] \exp[ i 2(q-q_2)\fdia p_1 \nonumber \\
& + i 2(q_1-q)\fdia p_2 + i 2(q_2-q_1)\fdia p]\ \mathcal{D}[p_1,p_2,q_1,q_2] ,
\end{eqnarray}
where $\mathcal{N}_0=2^\Omega$. Here we used the transformation of Grassman integration variables:
\begin{equation}
\int W[\theta] \mathcal{D}[\theta] = \int W[c\theta'] \frac{1}{c} \mathcal{D}[\theta'] ,
\label{gvartra}
\end{equation}
where $c$ is an arbitrary complex c-number constant.


\section*{References}


\begin{thebibliography}{10}

\bibitem{nc}
M.~A. Nielsen and I.~L. Chuang.
\newblock {\em Quantum Computation and Quantum Information}.
\newblock Cambridge University Press, Cambridge, England, 2000.

\bibitem{mw}
L.~Mandel and E.~Wolf.
\newblock {\em Optical coherence and quantum optics}.
\newblock Cambridge University Press, New York, 1995.

\bibitem{contvar1}
S.~L. Braunstein and P. Van~Loock.
\newblock Quantum information with continuous variables.
\newblock {\em Rev. Mod. Phys.}, 77:513, 2005.

\bibitem{weedbrook}
C. Weedbrook, S. Pirandola, R. Garc{\'\i}a-Patr{\'o}n, N.~J. Cerf, T.~C. Ralph, J.~H. Shapiro, and S. Lloyd.
\newblock Gaussian quantum information.
\newblock {\em Rev. Mod. Phys.}, 84:621--669, 2012.

\bibitem{contvar2}
G. Adesso, S. Ragy, and A.~R. Lee.
\newblock Continuous variable quantum information: Gaussian states and beyond.
\newblock {\em Open Syst. Inf. Dyn.}, 21:1440001, 2014.

\bibitem{groenewold}
H.~J. Groenewold.
\newblock On the principles of elementary quantum mechanics.
\newblock {\em Physica}, 12:405--460, 1946.

\bibitem{moyal}
J.~E. Moyal.
\newblock Quantum mechanics as a statistical theory.
\newblock {\em Math. Proc. Camb. Philos. Soc.}, 45:99--124, 1949.

\bibitem{psqm}
T.~L. Curtright and C.~K. Zachos.
\newblock Quantum mechanics in phase space.
\newblock {\em Asia Pacific Physics Newsletter}, 1:37--46, 2012.

\bibitem{wigner}
E. Wigner.
\newblock On the quantum correction for thermodynamic equilibrium.
\newblock {\em Phys. Rev.}, 40:749--759, 1932.

\bibitem{husimi}
K. Husimi.
\newblock Some formal properties of the density matrix.
\newblock {\em Nippon Sugaku-Buturigakkwai Kizi Dai 3 Ki}, 22:264--314, 1940.

\bibitem{sudarshan}
E.~C.~G. Sudarshan.
\newblock Equivalence of semiclassical and quantum mechanical descriptions of
  statistical light beams.
\newblock {\em Phys. Rev. Lett.}, 10:277--279, 1963.

\bibitem{glauber}
R.~J. Glauber.
\newblock Coherent and incoherent states of the radiation field.
\newblock {\em Phys. Rev.}, 131:2766--2788, 1963.

\bibitem{bastiaans1}
M.~J. Bastiaans.
\newblock The {W}igner distribution function applied to optical signals and
  systems.
\newblock {\em Opt. Commun.}, 25:26--30, 1978.

\bibitem{varilly}
J.~C. V{\'a}rilly and J.~M. Gracia-Bond{\'i}a.
\newblock The {M}oyal representation for spin.
\newblock {\em Annals of physics}, 190(1):107--148, 1989.

\bibitem{brif}
C.~Brif and A.~Mann.
\newblock Phase-space formulation of quantum mechanics and quantum-state
  reconstruction for physical systems with {L}ie-group symmetries.
\newblock {\em Phys. Rev. A}, 59:971--987, 1999.

\bibitem{mrowc}
S.~Mrowczynski and B.~Mueller.
\newblock Wigner functional approach to quantum field dynamics.
\newblock {\em Phys. Rev. D}, 50:7542--7552, 1994.

\bibitem{stquad}
F.~S. Roux.
\newblock Combining spatiotemporal and particle-number degrees of freedom.
\newblock {\em Phys. Rev. A}, 98:043841, 2018.

\bibitem{berra}
J. Berra-Montiel and A. Molgado.
\newblock Coherent representation of fields and deformation quantization.
\newblock {\em International Journal of Geometric Methods in Modern Physics},
  17:2050166, 2020.

\bibitem{caglaub}
K.~E. Cahill and R.~J. Glauber.
\newblock Density operators for fermions.
\newblock {\em Phys. Rev. A}, 59:1538, 1999.

\bibitem{fermipfunc}
J.~F. Corney and P.~D. Drummond.
\newblock Gaussian phase space representations for fermions.
\newblock {\em Phys. Rev. B}, 73:125112, 2006.

\bibitem{fermibarnett}
B.~J. Dalton, J.~Jeffers, and S.~M. Barnett.
\newblock Grassmann phase space methods for fermions: {I}. {M}ode theory.
\newblock {\em Annals of Physics}, 370:12--66, 2016.

\bibitem{fermibarnett2}
B.~J. Dalton, J.~Jeffers, and S.~M. Barnett.
\newblock Grassmann phase space methods for fermions. {II}. {F}ield theory.
\newblock {\em Annals of Physics}, 377:268--310, 2017.

\bibitem{polyakov}
E.~A. Polyakov.
\newblock Grassmann phase-space methods for fermions: {U}ncovering the
  classical probability structure.
\newblock {\em Phys. Rev. A}, 94:062104, 2016.

\bibitem{fermiqfunc}
R.~R. Joseph, L.~E.~C. Rosales-Z{\'a}rate, and P.~D. Drummond.
\newblock Phase space methods for {M}ajorana fermions.
\newblock {\em J. Phys. A: Math. Gen.}, 51:245302, 2018.

\bibitem{fermiwig}
S. Mr{\'o}wczy{\'n}ski.
\newblock Wigner functional of fermionic fields.
\newblock {\em Phys. Rev. D}, 87:065026, 2013.

\bibitem{araki}
H. Araki.
\newblock On the diagonalization of a dilinear {H}amiltonian by a {B}ogoliubov
  transformation.
\newblock {\em Publ. RIMS, Kyoto Univ. Ser. A}, 4:387--412, 1968.

\bibitem{ruijsenaars1}
S.~N.~M. Ruijsenaars.
\newblock On {B}ogoliubov transformations for systems of relativistic charged
  particles.
\newblock {\em J. Math. Phys.}, 18:517--526, 1977.

\bibitem{ruijsenaars2}
S.~N.~M. Ruijsenaars.
\newblock On {B}ogoliubov transformations. {II}. {T}he general case.
\newblock {\em Ann. Phys.}, 116:105--134, 1978.

\bibitem{svozil}
K.~Svozil.
\newblock Squeezed fermion states.
\newblock {\em Phys. Rev. Lett.}, 65:3341, 1990.

\bibitem{goldstein9}
H.~Goldstein.
\newblock {\em Classical Mechanics, 2nd Ed.}
\newblock Addison-Wesley Publishing Company, Reading, Massachusetts, USA, 1980.

\end{thebibliography}

\end{document}